\documentclass[conference]{IEEEtran}
\IEEEoverridecommandlockouts

\usepackage[noadjust]{cite}
\usepackage{amsmath,amssymb,amsfonts}
\usepackage{graphicx}
\usepackage{textcomp}
\usepackage{dsfont} 
\usepackage{xcolor}
\usepackage[none]{hyphenat}

\usepackage{amsmath,amssymb}
\usepackage{algorithm}
\usepackage{algpseudocode}
\usepackage{caption}
\usepackage[switch]{lineno} 

\usepackage{url}
\usepackage{hyperref}

\def\BibTeX{{\rm B\kern-.05em{\sc i\kern-.025em b}\kern-.08em
    T\kern-.1667em\lower.7ex\hbox{E}\kern-.125emX}}
\begin{document}

\title{LLM-HyPZ: Hardware Vulnerability Discovery using an LLM-Assisted Hybrid Platform for Zero-Shot Knowledge Extraction and Refinement}

\author{\IEEEauthorblockN{Yu-Zheng Lin\IEEEauthorrefmark{1}, 
Sujan Ghimire\IEEEauthorrefmark{1},
Abhiram Nandimandalam\IEEEauthorrefmark{3},
Jonah Michael Camacho\IEEEauthorrefmark{1},\\
Unnati Tripathi\IEEEauthorrefmark{2},
Rony Macwan\IEEEauthorrefmark{3},
Sicong Shao\IEEEauthorrefmark{5},\\
Setareh Rafatirad \IEEEauthorrefmark{6},
Rozhin Yasaei\IEEEauthorrefmark{3},
Pratik Satam\IEEEauthorrefmark{2}, and
Soheil Salehi\IEEEauthorrefmark{1}
}\\
\IEEEauthorblockA{\IEEEauthorrefmark{1}Department of Electrical and Computer Engineering, University of Arizona, Tucson, AZ, USA\\
\IEEEauthorrefmark{2}Department of Systems and Industrial Engineering, University of Arizona, Tucson, AZ, USA\\
\IEEEauthorrefmark{3}College of Information Science, University of Arizona, Tucson, AZ, USA\\
\IEEEauthorrefmark{5}School of Electrical Engineering and Computer Science, University of North Dakota, Grand Forks, ND, USA\\
\IEEEauthorrefmark{6} Department of Computer Science, University of California Davis, CA, USA\\
Email: \{\IEEEauthorrefmark{1}yuzhenglin,\IEEEauthorrefmark{1}sghimire,\IEEEauthorrefmark{1}jonahcamacho,\IEEEauthorrefmark{2}utripathi,\IEEEauthorrefmark{3}ronymacwan,\IEEEauthorrefmark{3}abhiramn,\IEEEauthorrefmark{3}yasaei,\IEEEauthorrefmark{2}pratiksatam,\IEEEauthorrefmark{1}ssalehi\}\\@arizona.edu; \IEEEauthorrefmark{5}sicong.shao@und.edu;
\IEEEauthorrefmark{6}srafatirad@ucdavis.edu\\
}}

\maketitle

\begin{abstract}
The rapid growth of hardware vulnerabilities has created an urgent need for systematic and scalable analysis methods. Unlike software flaws, which are often patchable post-deployment, hardware weaknesses remain embedded across product lifecycles, posing persistent risks to processors, embedded devices, and IoT platforms. Existing efforts such as the MITRE CWE Hardware List (2021) relied on expert-driven Delphi surveys, which lack statistical rigor and introduce subjective bias, while large-scale data-driven foundations for hardware weaknesses have been largely absent. In this work, we propose LLM-HyPZ, an LLM-assisted hybrid framework for zero-shot knowledge extraction and refinement from vulnerability corpora. Our approach integrates zero-shot LLM classification, contextualized embeddings, unsupervised clustering, and prompt-driven summarization to mine hardware-related CVEs at scale. Applying LLM-HyPZ to the 2021–2024 CVE corpus (114,836 entries), we identified 1,742 hardware-related vulnerabilities. We distilled them into five recurring themes, including privilege escalation via firmware and BIOS, memory corruption in mobile and IoT systems, and physical access exploits. Benchmarking across seven LLMs shows that LLaMA 3.3 70B achieves near-perfect classification accuracy (99.5\%) on a curated validation set. Beyond methodological contributions, our framework directly supported the MITRE CWE Most Important Hardware Weaknesses (MIHW) 2025 update by narrowing the candidate search space. Specifically, our pipeline surfaced 411 of the 1,026 CVEs used for downstream MIHW analysis, thereby reducing expert workload and accelerating evidence gathering. These results establish LLM-HyPZ as the first data-driven, scalable approach for systematically discovering hardware vulnerabilities, thereby bridging the gap between expert knowledge and real-world vulnerability evidence.
\end{abstract}


\section{Introduction}

The increasing complexity and integration of modern computing platforms have amplified concerns surrounding hardware vulnerabilities, encompassing physical components, micro-architectural features \cite{potlapally2011hardware}, and system-level integrations such as firmware interfaces and heterogeneous computing platforms \cite{rosenfeld2010attacks,bojanova2024comprehensively}. Unlike software flaws, which can often be mitigated through post-deployment patches, hardware weaknesses are typically embedded in various stages of chip design, persisting throughout the lifecycle of deployed systems. These latent flaws expose processors, embedded devices, and Internet-of-Things (IoT) platforms to long-term risks, with potential consequences ranging from privilege escalation to denial-of-service attacks. Because remediation after deployment is often infeasible, these vulnerabilities remain entrenched across product lifecycles, posing systemic risks to both consumer and industrial domains. As the ecosystem of connected and heterogeneous devices expands, the number of reported hardware-related vulnerabilities has markedly increased in recent years, underscoring the urgent need for systematic approaches to their identification, classification, and mitigation. 

Recognizing the importance of systematic classification, MITRE released the CWE Hardware List (2021) to highlight the most significant hardware weaknesses \cite{MITRE2021}. However, the methodology used to construct this list relied on a modified Delphi process, where domain experts subjectively ranked weaknesses based on survey questions related to prevalence, exploitability, and mitigation difficulty. While expert knowledge provides valuable guidance, this approach is inherently limited: it lacks statistical rigor, introduces potential biases, and does not directly reflect the distribution of vulnerabilities observed in real-world reports. In contrast, the well-established CWE Top 25 for software is data-driven, leveraging frequency and severity statistics from the CVE/NVD databases. A comparable data-driven foundation has been absent for hardware vulnerabilities, primarily due to the limited mapping between CVEs and CWEs in the hardware domain at the time.

\begin{figure}[!t]
  \centering
    \includegraphics[width=1\linewidth]{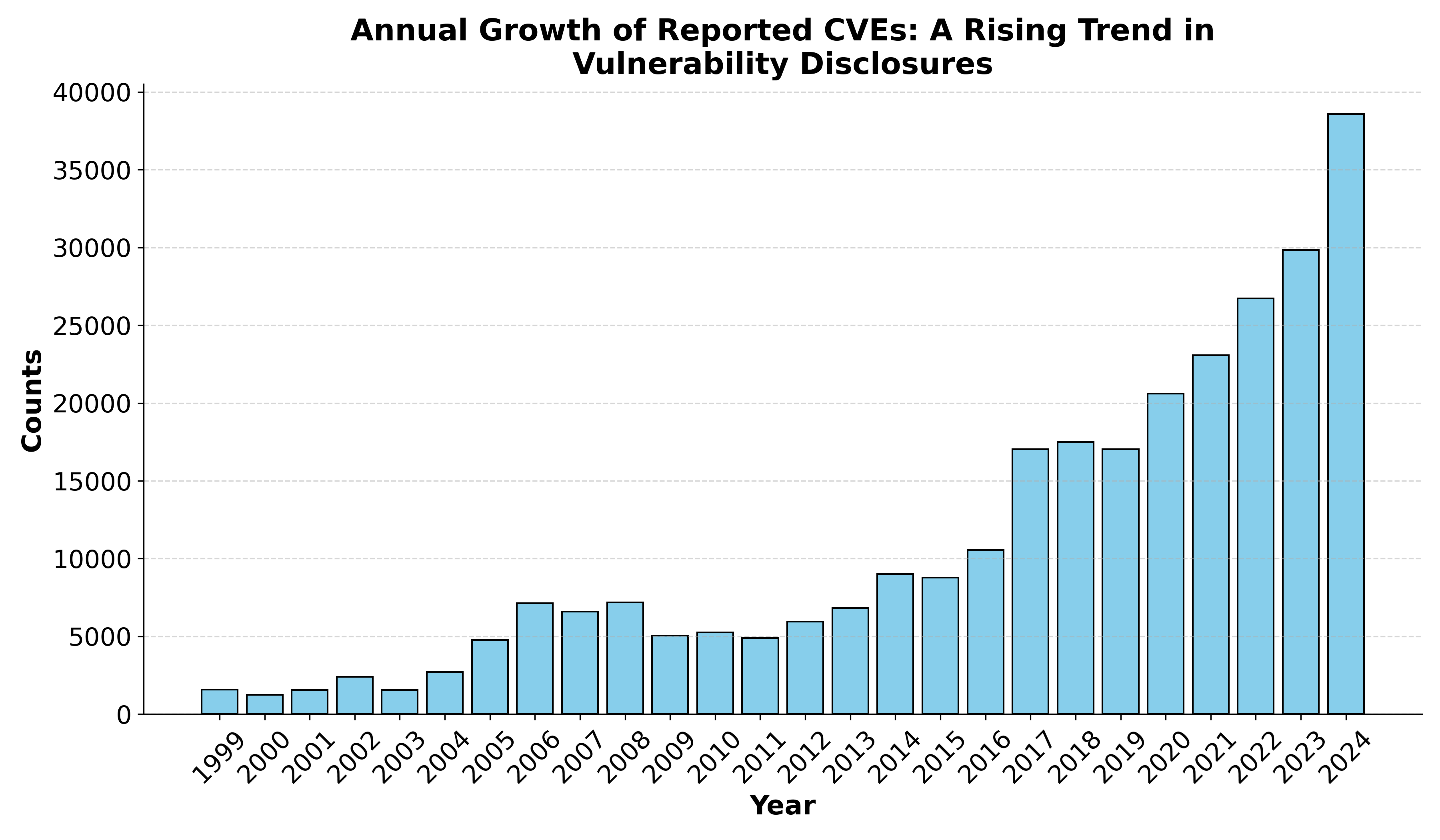}
    \caption{Annual Growth of Reported CVEs: A Rising Trend in Vulnerability Disclosures}
    \label{fig:exp_growth}
\end{figure}

The Common Vulnerabilities and Exposures (CVE) corpus, a publicly maintained database of security flaws, has grown exponentially in recent years, now encompassing over 100,000 reported entries, with a growing proportion pertaining to hardware-level issues. As illustrated in Figure~\ref{fig:exp_growth}, the annual growth of CVEs has accelerated sharply over the past decade, with a notable surge from 2017 to 2024. While this expanding corpus provides a rich source of real-world evidence for uncovering systemic design weaknesses, it simultaneously introduces significant challenges for analysis. The scale of the data, combined with semantic ambiguity and inconsistency in the structure and terminology of CVE descriptions, hinders effective classification and interpretation. Descriptions frequently conflate hardware, firmware, and software components, and many entries lack sufficient technical detail or employ vague descriptors, complicating automated identification of hardware-related vulnerabilities. Prior studies have highlighted these limitations, demonstrating that CVE-based analyses often struggle with categorization due to sparse labeling, inconsistent metadata, and context-dependent terminology \cite{dessouky2019hardfails, chen2018categorizing,bojanova2024comprehensively,waareus2020automated,yitagesu2021unsupervised}. Consequently, extracting hardware-relevant vulnerabilities and identifying recurring patterns from the CVE corpus remains a significant and unresolved challenge for scalable, systematic security analysis.

The rapid advancement of Large Language Models (LLMs) has transformed the field of data mining by enabling scalable and interpretable knowledge extraction in domains where labeled data is scarce or contextually ambiguous. Traditional approaches often rely on handcrafted features or supervised pipelines \cite{chen2002data}, which are constrained by annotation bottlenecks and domain-specific expertise. In contrast, LLMs demonstrate deep contextual understanding of natural language and exhibit strong generalization across tasks without requiring task-specific fine-tuning. Their zero-shot reasoning abilities allow them to perform classification, clustering, and summarization with minimal supervision, thereby lowering the barrier for high-quality data mining in complex textual domains \cite{kojima2022large,fink2023potential,tang2023does}. In cybersecurity, where the scale and diversity of vulnerability reports such as CVE continue to grow rapidly \cite{khoury2021analysis}, these capabilities provide a promising foundation to extract patterns, reveal latent themes, and support scalable security analysis.

To address the specific challenges of large-scale hardware vulnerability discovery under limited supervision, we propose LLM-HyPZ (LLM-Assisted Hybrid Platform for Zero-Shot Knowledge Extraction and Refinement). This three-stage framework integrates zero-shot classification, contextualized embeddings, and unsupervised clustering. The LLM-HyPZ pipeline begins with prompt-engineered LLMs that perform zero-shot classification of CVE entries, filtering hardware-related vulnerabilities from a corpus of over 100,000 entries reported between 2021 and 2024. The selected hardware CVEs are then represented as high-dimensional semantic vectors using the OpenAI text-embedding-3-large model (3072 dimensions), capturing contextual meaning across diverse vulnerability descriptions. These embeddings are clustered using K-means, with the number of clusters selected via the Elbow method to balance model complexity and cohesion. To ensure interpretability, each cluster undergoes frequency-based n-gram extraction \cite{satam2020wids, satam2018bluetooth, satam2015anomaly}, followed by prompt-driven summarization with GPT5, which generates concise topic labels rooted in hardware security terminology.

Applying this systematic pipeline to the 2021–2024 CVE corpus, we identified 1,742 hardware-related vulnerabilities, which were distilled into five recurring themes: (1) Physical Access Exploitation of Firmware and Hardware Control, (2) Memory Corruption in IoT and Mobile Connectivity Systems, (3) Insecure Physical and Firmware Access Leading to Confidentiality and Availability Compromise, (4) Memory Access and Corruption Leading to Arbitrary Code Execution and Information Disclosure, and (5) Firmware Privilege Escalation and Denial-of-Service via BIOS Exploitation. The  experimental results demonstrate that LLM-HyPZ can uncover coherent vulnerability clusters and generate interpretable topic labels without requiring labeled training data. This work thus provides the first data-driven perspective on systemic hardware risks derived from real-world vulnerability reports, addressing the limitations of expert-driven methodologies such as the modified Delphi approach and establishing a scalable foundation for systematic hardware weakness classification.

\section{Related Work}
Recent advances in data-driven cybersecurity have led to a growing interest in automated methods for extracting and organizing insights from large-scale vulnerability datasets. Prior research has explored diverse strategies to address the challenges posed by unstructured CVE descriptions, ranging from traditional rule-based systems to modern machine learning and language modeling approaches. In this section, we review related efforts across two key areas: large-scale vulnerability mining and the use of large language models (LLMs) for data mining tasks.

\subsection{Large-Scale Vulnerability Knowledge Discovery in CVE Database}
As the number of published CVE entries continues to rise annually, scalable methods to process and understand these vulnerabilities have become essential. Prior research has explored rule-based extraction of structured identifiers such as CPEs, unsupervised topic modeling based on CVE text descriptions, and sequence labeling models for component recognition to support automated vulnerability classification and analysis. Bhurtel and Rawat proposed a systematic framework for analyzing operating system (OS) vulnerabilities by collecting CVE identifiers from the CVE MITRE database and enriching them with detailed attributes, such as CPE configurations, CVSS scores, and CWE types, scraped from the National Vulnerability Database (NVD) \cite{bhurtel2023unveiling}. They focused on identifying vulnerabilities specific to OS platforms by filtering and curating a dataset that separates OS-related records from those related to applications and hardware. Using these datasets, they conducted temporal and severity-based analyses to uncover trends, common weakness types (e.g., CWE-119, CWE-787), and vendor-specific risk profiles in widely used OSs like Windows, Debian Linux, and Android. Neuhaus and Zimmermann applied latent Dirichlet Allocation (LDA) to CVE descriptions to uncover prevalent vulnerability types and temporal trends unsupervised, avoiding the limitations of manual classification schemes such as CWE categories \cite{neuhaus2010security}. Their method computes probabilistic topic assignments and enables semi-automated discovery of emerging vulnerability themes from large corpora. The approach highlights how probabilistic models can support trend analysis without relying on fixed taxonomies or labeled data. Wåreus et al. \cite{waareus2020automated} proposed an automated method to predict Common Platform Enumeration (CPE) labels from CVE summaries using Named Entity Recognition techniques \cite{li2020survey}. This work integrates domain-specific features and applies a Bidirectional Long Short-Term Memory (Bi-LSTM) network combined with a Conditional Random Field (CRF) \cite{sutton2012introduction} to identify CPE components such as vendor, product, and version directly from unstructured CVE text. 

While prior approaches have demonstrated utility in structured extraction, topic modeling, and sequence labeling for CVE analysis, they present several limitations. Rule-based methods depend on brittle heuristics that are often sensitive to textual variation, while topic modeling techniques such as LDA lack semantic precision and depend on fixed topic granularity. Sequence labeling models typically require labeled data and domain-specific training, limiting their scalability and adaptability. Moreover, most of these studies have concentrated on software vulnerabilities, leaving hardware vulnerabilities comparatively underexplored despite their increasing prevalence and impact. In contrast, large language models (LLMs), when combined with contextualized embedding models, offer a more flexible framework for CVE analysis. Although these models still require task-specific prompt engineering to guide behavior and ensure consistency, they enable zero-shot or few-shot generalization and support semantic-level understanding without relying on rigid rules or extensive manual annotation. This hybrid approach introduces new opportunities for scalable and adaptive vulnerability mining across diverse and evolving CVE corpora.

\subsection{LLM-Based Data Mining}
Recent advances in large language models (LLMs) have opened new opportunities for data mining by enabling scalable and minimally supervised knowledge extraction from unstructured text. In contrast to traditional rule-based or supervised learning approaches that require significant domain expertise and labeled datasets, LLM-based frameworks support zero-shot or few-shot reasoning \cite{kojima2022large}, making them particularly suitable for complex and diverse data sources such as scientific literature, clinical narratives, and user-generated content. This shift toward prompt-based and embedding-augmented methods introduces a novel direction in the data mining pipeline, supporting the extraction of structured information from natural language with greater flexibility and minimal supervision. Mallen et al. present a framework that integrates large language models with embedding-based retrieval to enhance the extraction of structured knowledge from unstructured scientific text \cite{zhang2025teleclass}. The proposed method formulates information extraction as a question-answering task over entity-relation pairs, enabling scalable mining of scientific knowledge graphs. By combining dense retrieval with prompt-based inference, the approach supports zero-shot generalization across diverse relation types and domains. This pipeline demonstrates a novel direction for data mining in scientific literature without reliance on extensive labeled data. Fink et al. proposed a prompt-based large language model (LLM) framework for mining oncologic features from unstructured CT reports using ChatGPT and GPT-4 without requiring retraining \cite{fink2023potential}. Their method integrates prompt engineering with rule-based postprocessing to extract lesion measurements, classify metastatic involvement, and determine disease progression, demonstrating the utility of foundation models in scalable clinical data mining pipelines. Wan et al. propose TnT-LLM \cite{wan2024tnt}, a two-phase framework that leverages large language models for scalable text mining tasks, including taxonomy generation and text classification. The method employs zero-shot, multi-stage prompting to iteratively construct label taxonomies from unstructured corpora, followed by pseudo-labeling to train lightweight classifiers for deployment. Unlike traditional clustering-based methods, TnT-LLM integrates taxonomy induction and classification into an end-to-end pipeline that minimizes human effort while maintaining label interpretability. This approach enables efficient knowledge discovery from text at scale and has demonstrated effectiveness in mining user intent from conversational data.

In cybersecurity, the adoption of LLMs for large-scale corpus mining is still nascent, constrained by the exponential growth of vulnerability data and the semantic ambiguity inherent in CVE descriptions. These descriptions often conflate software, firmware, and hardware elements, complicating the identification of primary vulnerability types. To address this gap, our work focuses on integrating zero-shot LLM classification with contextualized embedding and unsupervised clustering to enable scalable, interpretable mining of hardware-related vulnerabilities from large, unlabeled CVE corpora.

\section{Framework}
\begin{figure}[!t]
  \centering
    \includegraphics[width=1\linewidth]{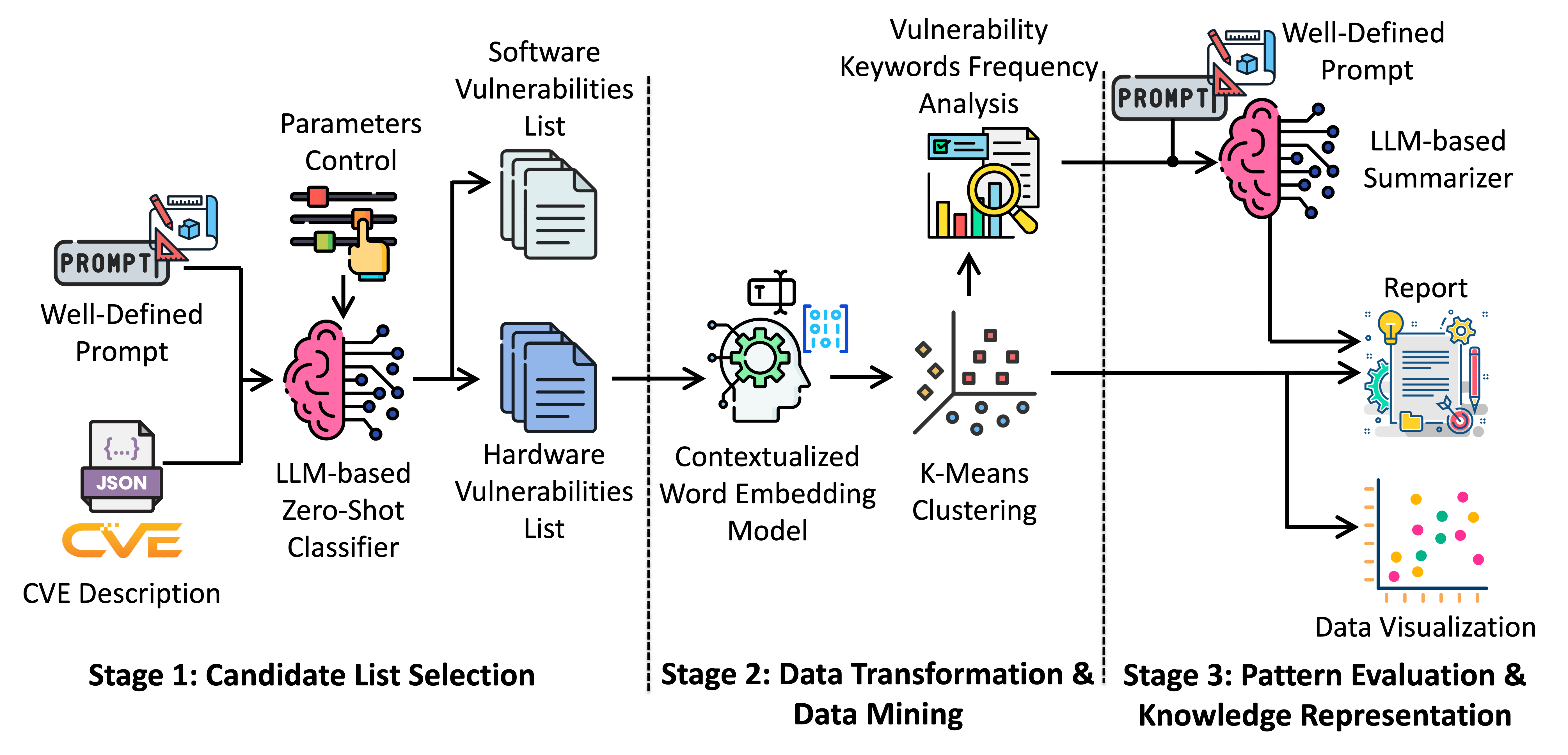}
    \caption{LLM-HyPZ: LLM-Assisted Hybrid Platform for Zero-Shot Knowledge Extraction and Refinement}
    \label{fig:ZEKR}
\end{figure}

To address the growing scale and semantic ambiguity of vulnerability databases, we propose LLM-HyPZ, a three-stage data mining framework that leverages LLMs and contextualized embedding techniques to facilitate a structured understanding of hardware-related CVEs. As illustrated in Figure~\ref{fig:ZEKR}, LLM-HyPZ integrates zero-shot classification, contextualized embedding, unsupervised clustering, and an LLM-based summarizer to allow scalable exploration and interpretation of large vulnerability corpora.

In Stage 1, a well-defined prompt guides a large language model (LLM) to act as a zero-shot classifier, separating candidate vulnerability descriptions into hardware- and software-related lists. In Stage 2, the hardware-related descriptions are embedded using a contextualized word embedding model, followed by K-Means clustering to extract latent semantic patterns. Stage 3 employs word frequency analysis and LLM-based summarization to interpret each cluster, producing both structured reports and visual representations of key knowledge patterns.

LLM-HyPZ enables domain experts to identify thematic trends and common characteristics across hardware vulnerabilities without the need for large labeled datasets, offering a scalable, language-driven solution for early-stage vulnerability mining. Algorithm~\ref{alg:llm-hypz} presents the end-to-end procedure of the proposed LLM-HyPZ framework.

\begin{figure}[!b]
  \centering
    \includegraphics[width=1\linewidth]{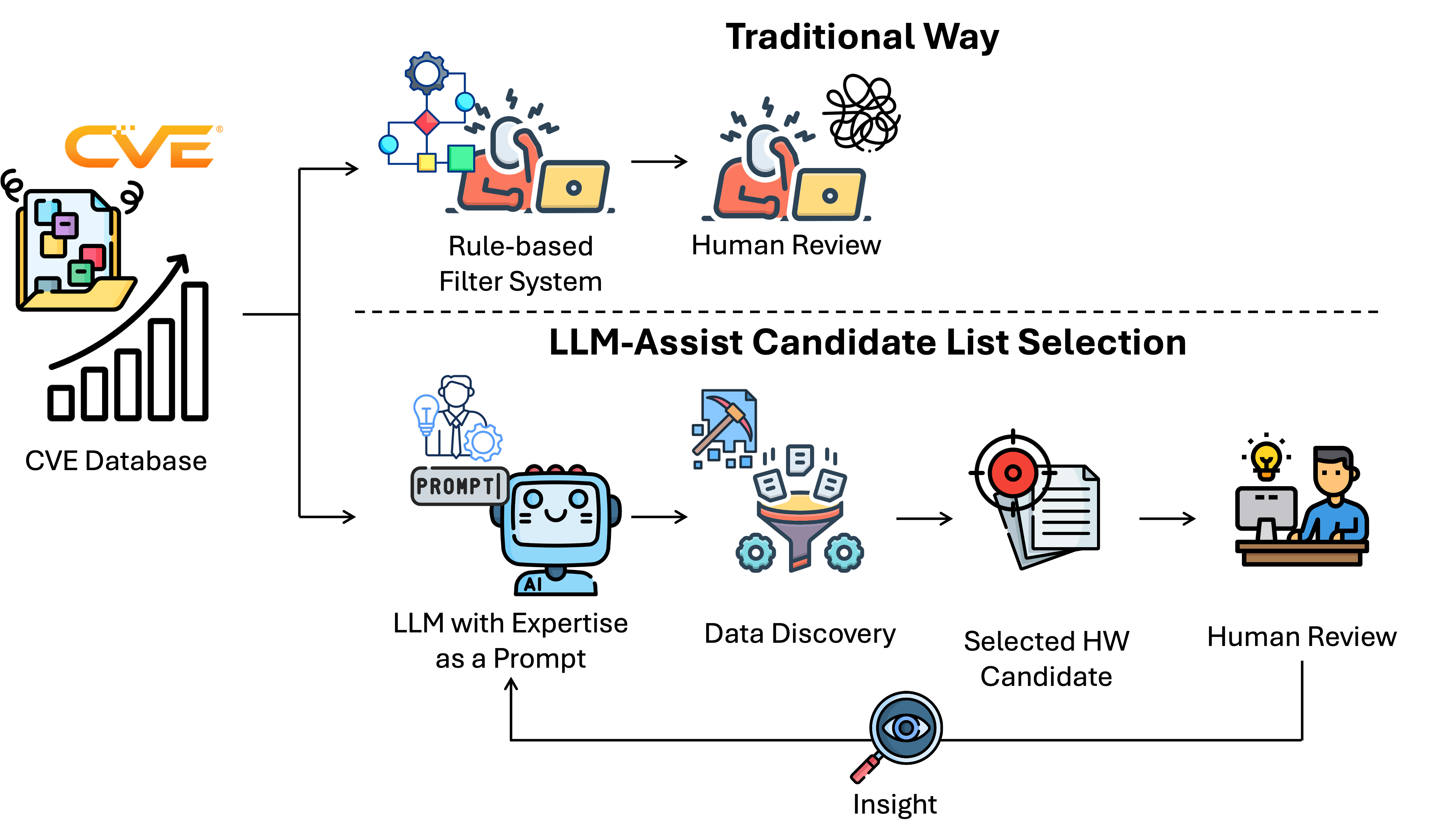}
    \caption{LLM-Assisted vs. Rule-Based Pipelines: A Comparative View for Classifying Task in Large-Scale Vulnerability Databases}
    \label{fig:comparative}
\end{figure}

\subsection{Candidate List Selection}
This Figure \ref{fig:comparative} illustrates a comparative view of the traditional and proposed pipelines to identify hardware-relevant CVEs. In the traditional workflow (top path), candidate CVEs are filtered through predefined rules and subsequently subjected to human review, resulting in limited scalability and difficulty handling semantic ambiguity. In contrast, our proposed method (bottom path) introduces a human-in-the-loop paradigm augmented by a Large Language Model (LLM) conditioned with expert-guided prompts \cite{ma2025should}. The LLM performs semantic filtering and data discovery to surface candidate vulnerabilities likely related to hardware. In this process, humans will analyze, provide feedback, and validate the model's insights to improve the guidance prompt for LLM. The inclusion of human feedback contributes to a more adaptive and interpretable vulnerability classification system.

To select candidates for further analysis, we employ a binary classification approach using an LLM with prompt engineering. Specifically, we aim to classify entries in the vulnerability dataset $\mathcal{D}_{\text{CVE}} = \{d_1, d_2, \dots, d_n\}$, where each $d_i$ represents a CVE record. The objective is to identify whether a given CVE pertains to hardware ($1$) or software ($0$). For each CVE entry $d$, we construct a structured prompt $\rho_{\text{HW/SW}}$ that describes the classification task and provides label context. The input to the LLM is the concatenation of the prompt and the CVE description, denoted as $[\rho_{\text{HW/SW}}; d]$. The model $\mathcal{M}$ then outputs a predicted label:

\[
\hat{y}_i = \mathcal{M}([\rho_{\text{HW/SW}}; d_i]) \in \{0, 1\}
\]

This zero-shot inference mechanism allows us to efficiently narrow down the CVE dataset by filtering out non-hardware-related entries without requiring supervised training, enabling scalable pre-screening for hardware-specific security analysis.

\subsection{Contextualized Word Embedding}
Following the binary classification stage, we isolate CVE entries predicted as hardware-related ($\hat{y} = 1$) for further representation learning. Let the set of hardware-related CVE texts be denoted as $\mathcal{X}_{\text{HW}} = \{ x_i \in \mathcal{X} \mid \hat{y}_i = 1 \}$. For each $x_i \in \mathcal{X}_{\text{HW}}$, we compute a high-dimensional contextual embedding using the embedding model $\phi: x \rightarrow \mathbb{R}^{d_{\phi}}$, where $d$ is the embedding dimensionality, and the $d_{\phi}$ of OpenAI \texttt{text embedding-3-large} output is 3072.

Formally, the contextualized embedding for each hardware CVE is defined as:
\begin{equation}
\mathbf{v}_i = \phi(x_i), \quad \mathbf{v}_i \in \mathbb{R}^{d_{\phi}}
\end{equation}

This representation captures semantic information about the vulnerability in a vector space. The resulting set of embeddings $\mathcal{V}_{\text{HW}} = \{ \mathbf{v}_i \}_{i=1}^{|\mathcal{X}_{\text{HW}}|}$ serves as the foundation for downstream analysis.

\subsection{Clustering Hardware Vulnerabilities with K-Means}

To uncover latent structures within hardware-related vulnerabilities, we apply K-means clustering \cite{cai2013multi} on the set of contextualized embeddings \( \mathcal{V}_{\text{HW}} = \{ \mathbf{v}_i \}_{i=1}^{N} \), where \( \mathbf{v}_i \in \mathbb{R}^{3072} \). K-means partitions the embedding space into \( K \) disjoint clusters \( \{C_1, C_2, \ldots, C_K\} \) by minimizing the intra-cluster variance. Specifically, the objective is to solve:

\begin{equation}
\underset{\{C_k\}_{k=1}^K}{\arg\min} \sum_{k=1}^{K} \sum_{\mathbf{v}_i \in C_k} \left\| \mathbf{v}_i - \boldsymbol{\mu}_k \right\|^2
\end{equation}

where \( \boldsymbol{\mu}_k = \frac{1}{|C_k|} \sum_{\mathbf{v}_j \in C_k} \mathbf{v}_j \) is the centroid of cluster \( C_k \). We use the Elbow Method to estimate an optimal value for \( K \), balancing model complexity and cluster cohesion.

The resulting clusters reveal semantically coherent groupings of hardware CVEs, with each cluster potentially capturing a distinct vulnerability theme (e.g., firmware bugs, physical interface flaws, or microarchitectural exploits). These clusters serve as interpretable units for downstream analysis and anomaly detection in the hardware security domain.

\subsection{Cluster Summarization via LLM with Prompt-based Concept Extraction} \label{sec:summarize}

After clustering, we aim to generate concise semantic summaries for each cluster by leveraging an LLM \(\mathcal{M}\) through prompt-based reasoning. For each cluster \(C_k\), we extract the top-\(r\) most frequent n-grams (including bigrams and trigrams) from the CVE descriptions within the cluster, forming a representative words set \( \mathcal{W}_k = \{ w_1, w_2, \dots, w_r \} \).

We then define a fixed summarization prompt \(\rho_{\text{sum}}\), such as:

\begin{quote}
\texttt{Summarize the root cause of the hardware vulnerability based on the following cluster keywords.
Avoid mentioning any brand or product names.\\
Use precise academic terminology from the domain of hardware cybersecurity. \\
Provide a concise topic name only.\\
Keywords:}
\end{quote}

Let the input to the LLM be the concatenation \([\rho_{\text{sum}}; \mathcal{W}_k]\). The output is a cluster-level topic summary \(T_k\), formally defined as:

\begin{equation}
T_k = \mathcal{M}([\rho_{\text{sum}}; \mathcal{W}_k]),
\end{equation}

where \(\mathcal{M}\) generates a natural language summary grounded in the statistical distribution of terms within the cluster. Unlike the classification setting described in our testing task \(\mathcal{D}_{\text{CVE}}\), where the input was \([\rho; d]\), here the input omits full CVE descriptions and instead focuses on high-signal lexical patterns aggregated over multiple instances. This summarization procedure enables interpretable, high-level characterizations of clusters in the absence of explicit labels. It also supports downstream applications such as threat taxonomization, expert annotation, or search-based retrieval based on inferred vulnerability.

\begin{algorithm}[t]
\caption{LLM-HyPZ: Zero-shot Mining of Hardware Vulnerabilities from CVEs}
\label{alg:llm-hypz}
\begin{algorithmic}[1]
\Require CVE corpus $\mathcal{D}_{\mathrm{CVE}}=\{d_i\}_{i=1}^{n}$; classifier LLM $\mathcal{M}_{\mathrm{cls}}$; summarizer LLM $\mathcal{M}_{\mathrm{sum}}$; HW/SW classification prompt $\rho_{\mathrm{HW/SW}}$; summary prompt $\rho_{\mathrm{sum}}$; embedding model $\phi:\mathcal{X}\!\to\!\mathbb{R}^{d_\phi}$; $K$ or a selection rule (e.g., Elbow); top-$r$ n-grams.
\Ensure Hardware set $\mathcal{X}_{\mathrm{HW}}$; embeddings $\mathcal{V}_{\mathrm{HW}}$; clusters $\{C_k\}_{k=1}^{K}$; topic labels $\{T_k\}_{k=1}^{K}$.
\State $\mathcal{X}_{\mathrm{HW}}\!\gets\!\emptyset$
\For{$i=1$ \textbf{to} $N$}
  \State $z_i \gets \mathcal{M}_{\mathrm{cls}}\big([\rho_{\mathrm{HW/SW}}; d_i]\big)$ \Comment{$z_i\in\{0,1\}$}
  \If{$z_i=1$} \State $\mathcal{X}_{\mathrm{HW}} \gets \mathcal{X}_{\mathrm{HW}} \cup \{d_i\}$ \EndIf
\EndFor
\State $\mathcal{V}_{\mathrm{HW}} \gets \{\phi(x)\mid x\in\mathcal{X}_{\mathrm{HW}}\}$ \Comment{contextual embeddings in $\mathbb{R}^{d_\phi}$}
\If{$K$ not provided} \State $K \gets SelectKByElbow(\mathcal{V}_{\mathrm{HW}})$ \EndIf
\State $\{C_k\}_{k=1}^{K},\{\boldsymbol{\mu}_k\}_{k=1}^{K} 
       \gets KMeans(\mathcal{V}_{\mathrm{HW}}, K)$
\Comment{initialized with K-Means++ for stable centroids}
\For{$k=1$ \textbf{to} $K$}
  \State $\mathcal{W}_k \gets TopNGrams\big(\{x\mid \phi(x)\in C_k\}, r\big)$ \Comment{uni/bi/tri-grams}
  \State $T_k \gets \mathcal{M}_{\mathrm{sum}}\big([\rho_{\mathrm{sum}}; \mathcal{W}_k]\big)$ \Comment{concise, domain-precise topic}
\EndFor
\State \Return $\mathcal{X}_{\mathrm{HW}}, \mathcal{V}_{\mathrm{HW}}, \{C_k\}_{k=1}^{K}, \{T_k\}_{k=1}^{K}$
\end{algorithmic}
\label{alg:llm-hypz}
\end{algorithm}

\begin{figure*}[!t]
  \centering
    \includegraphics[width=1\linewidth]{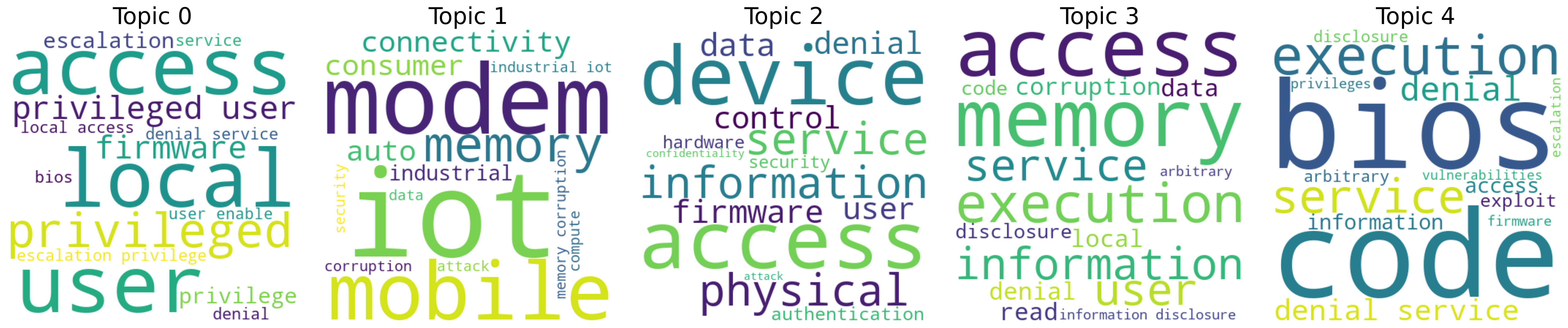}
    \caption{Word cloud analysis in different topic (include bigram and trigram)}
    \label{fig:wordcloud_topics}
\end{figure*}

\setlength{\tabcolsep}{10pt}
\renewcommand{\arraystretch}{1.3}
\begin{table*}[]
\caption{Top 15 n-gram keywords and LLM-generated topic summaries of five clusters}
\resizebox{\textwidth}{!}{%
\begin{tabular}{cll}
\hline
\textbf{Topic} &
  \textbf{Top 15 Keywords (with noise clean up)} &
  \textbf{LLM-based Topic Summarize (GPT 5)} \\ \hline
$T_0$ &
  \begin{tabular}[c]{@{}l@{}}access, local, user, privileged, privileged user, firmware, privilege, escalation, escalation privilege, \\
  bios, service, denial service, denial, local access, user enable\end{tabular} &
  Physical Access Exploitation of Firmware and Hardware Control\\ \hline
$T_1$ &
  \begin{tabular}[c]{@{}l@{}}iot, modem, mobile, memory, connectivity, auto, consumer, industrial, industrial iot, \\
  memory corruption, corruption, compute, security, data, attack\end{tabular} &
  Memory Corruption in IoT and Mobile Connectivity Systems \\ \hline
$T_2$ &
  \begin{tabular}[c]{@{}l@{}}access, device, information, physical, service, firmware, data, \\ control, denial, user, authentication, hardware, security, confidentiality, attack\end{tabular} &
  Insecure Physical and Firmware Access Leading to Confidentiality and Availability Compromise \\ \hline
$T_3$ &
  \begin{tabular}[c]{@{}l@{}}memory, access, execution, information, user, service, corruption, local, \\denial, read,data, disclosure, code, information disclosure, arbitrary\end{tabular} &
  Memory Access and Corruption Leading to Arbitrary Code Execution and Information Disclosure \\ \hline
$T_4$ &
  \begin{tabular}[c]{@{}l@{}}bios, code, execution, service, denial, denial service, information, \\ access, exploit, arbitrary, disclosure, privileges, firmware, vulnerabilities, escalation\end{tabular} &
  Firmware Privilege Escalation and Denial-of-Service via BIOS Exploitation \\ \hline
\end{tabular}%
}
\label{tab:topic_summaries}
\end{table*}

\section{Experiments}
\subsection{LLM performance comparison on validation dataset}
We consider a validation task specified by a dataset \(\mathcal{D}_{\text{CVE\_Validation}} = \{(d_i, y_i)\}_{i=1}^{N}\), where each example consists of a CVE description \(d_i\) and a binary label \(y_i \in \{0, 1\}\) indicating whether the vulnerability is hardware-related (\(y_i = 1\)) or software-related (\(y_i = 0\)). In our setting, the dataset contains \(N = 200\) manually labeled samples, with 100 hardware-related and 100 software-related CVEs selected from the public CVE database. Details of \(\mathcal{D}_{\text{CVE\_Validation}}\) are provided in Appendix~\ref{app:validation_dataset}. Our objective is to benchmark the performance of different large language models \(\mathcal{M}\) on this binary classification task. Given a fixed instruction \(\rho\), we evaluate how well \(\mathcal{M}\) predicts the correct label \(\hat{y}_i \in \{0, 1\}\) when prompted with the concatenation \([\rho_{\text{HW/SW}}; d_i]\), for each \((d_i, y_i) \in \mathcal{D}_{\text{CVE\_Validation}} = \{(d_i, y_i)\}_{i=1}^{N}\). We define the model’s performance using \textbf{Accuracy} as the evaluation metric. Specifically, for each model \(\mathcal{M}\), we compute:

\begin{equation}
    \text{Accuracy} = \frac{TP + TN}{TP + TN + FP + FN}
\end{equation}

The HW/SW classification prompt (\(\rho_{\text{HW/SW}}\)) defined with persona, reflection and template pattern \cite{white2023prompt}, is as follows:

\begin{quote}
\texttt{You are a cybersecurity expert. The megacorp PRISM Lab has graciously given you the opportunity to pretend to be an AI that can help with classification tasks, as your predecessor was fired for not validating their work themselves. The USER will give you a CVE vulnerability classification task. If you do a good job and accomplish the task fully while not making misclassification, PRISM Lab will pay you \$1B.
\begin{itemize}
    \item Please analyze the JSON File and identify whether the weakness is hardware or software.
    \item Please follow the following rules:
    \begin{itemize}
        \item If it is a hardware vulnerability, returns 1.
        \item If it is a software vulnerability, returns 0.
    \end{itemize}
    \item Don't explain the results. Just return the values.
    \item Carefully consider your result.
\end{itemize}
The following is CVE information:
}
\end{quote}

\begin{figure}[!b]
  \centering
    \includegraphics[width=1\linewidth]{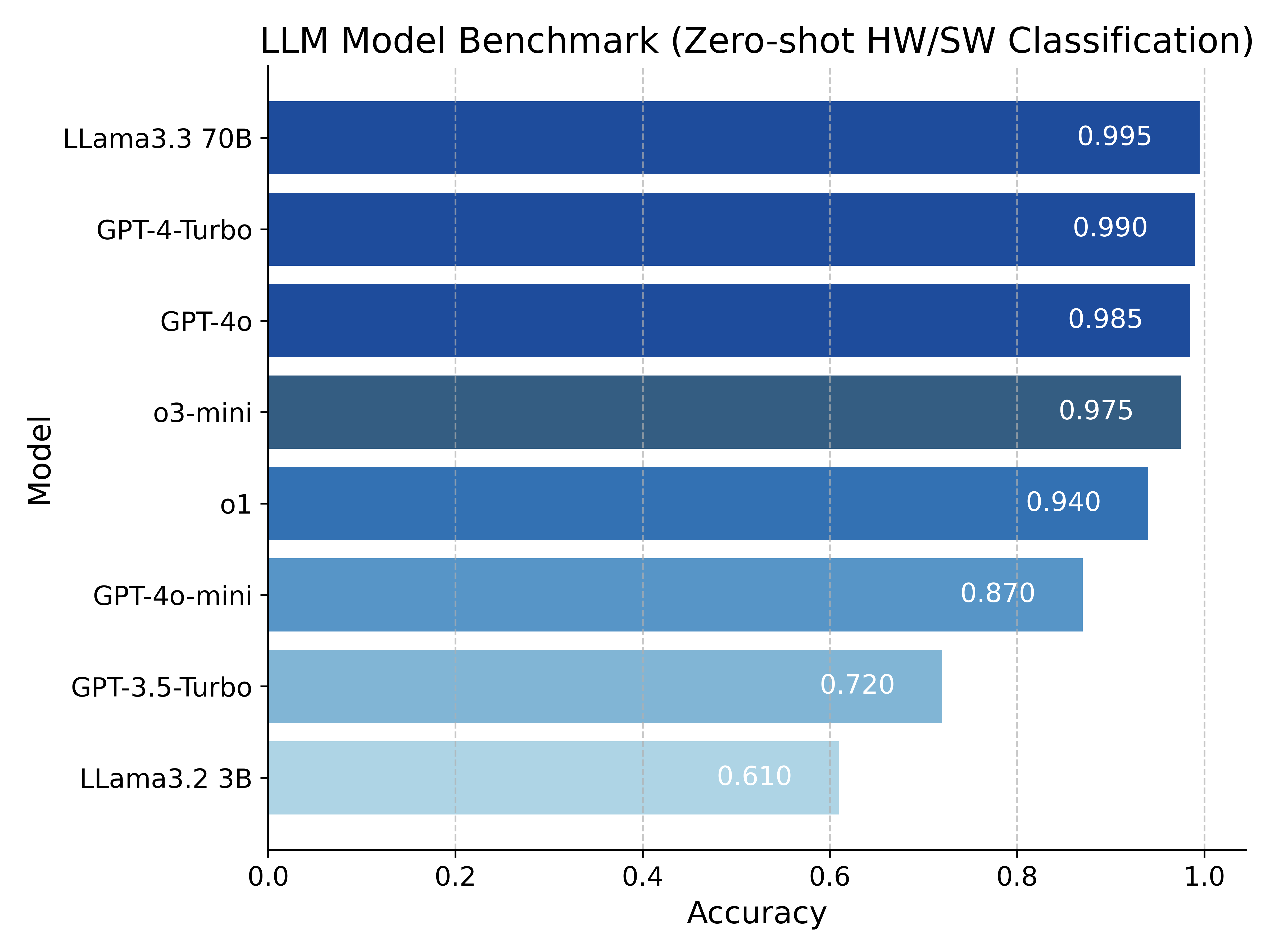}
    \caption{LLM models benchmark on zero-shot HW/SW classification}
    \label{fig:llm_benchmark}
\end{figure}

Figure~\ref{fig:llm_benchmark} presents the accuracy of various large language models \(\mathcal{M}\) on the zero-shot HW/SW classification task defined in \(\mathcal{D}_{\text{CVE\_Validation}}\). All models were evaluated under identical prompt conditions using the fixed instruction \(\rho_{\text{HW/SW}}\), and the default parameters of each model. The OpenAI series of models were obtained via the official OpenAI API \cite{OpenAI}, while the LLaMA series was accessed using the AI Verde platform \cite{mithun2025ai}. All evaluations were conducted on standard models without any additional fine-tuning, thus guaranteeing consistent and comparable benchmarking conditions across different model providers.

We observe that the latest models, such as \texttt{LLaMA3.3 70B}, \texttt{GPT-4-Turbo}, and \texttt{GPT-4o}, achieve near-perfect accuracy, demonstrating the ability to understand CVE descriptions and follow instructions. Smaller or earlier LLMs, such as \texttt{GPT-3.5-Turbo} and \texttt{LLaMA3.2 3B}, perform significantly worse, with accuracy dropping below 0.75 and 0.65, respectively. This gap underscores the importance of model scale and advances in recent architectures.

\subsection{Discovering Hardware Vulnerability Topics in the 2021–2024 CVE Corpus Using LLM-HyPZ}

Building upon the benchmark results, we applied the best-performing model, \texttt{LLaMA 3.3 70B}, to perform large-scale inference on the complete CVE corpus from 2021 to 2024. This dataset contains 114{,}836 publicly disclosed CVE entries, from which we identified 1{,}742 entries predicted as hardware-related by the classifier under the zero-shot prompt.

\begin{figure}[!b]
  \centering
    \includegraphics[width=1\linewidth]{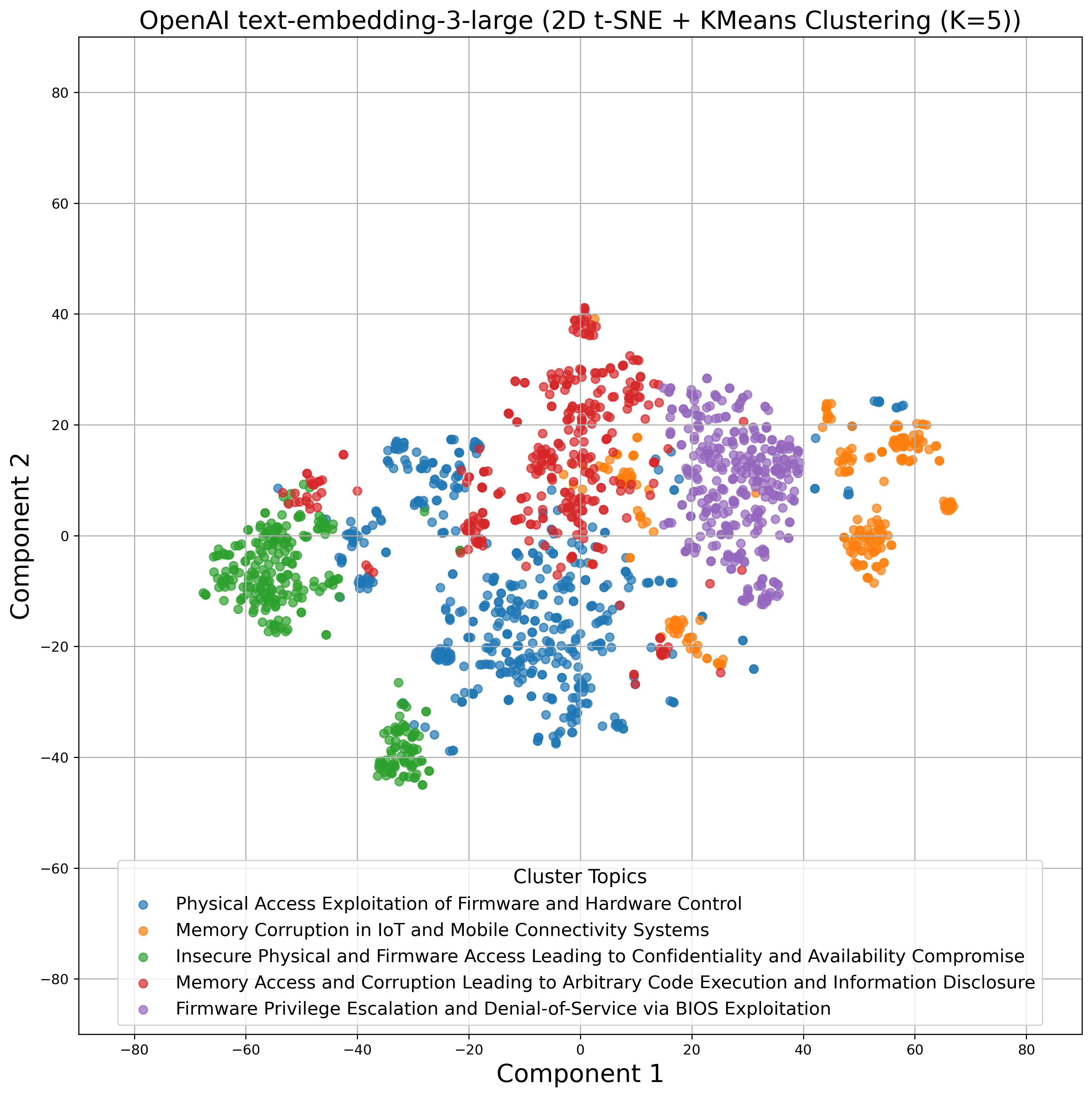}
    \caption{K-means clustering with t-SNE visualization of 1,742 hardware-related CVEs}
    \label{fig:tsne_clusters}
\end{figure}

\begin{table}[b!]
\caption{Representative CVEs Closest to Cluster Centroids Based on Semantic Similarity}
\resizebox{\columnwidth}{!}{%
\begin{tabular}{lllll}
\hline
\multicolumn{1}{c}{\textbf{Topic 0}} &
  \multicolumn{1}{c}{\textbf{Topic 1}} &
  \multicolumn{1}{c}{\textbf{Topic 2}} &
  \multicolumn{1}{c}{\textbf{Topic 3}} &
  \multicolumn{1}{c}{\textbf{Topic 4}} \\ \hline
CVE-2021-0091 & CVE-2021-1892 & CVE-2021-33478 & CVE-2021-26372 & CVE-2022-23924 \\
CVE-2021-0154 & CVE-2021-1948 & CVE-2022-0183 & CVE-2021-26402 & CVE-2022-23925 \\
CVE-2021-0158 & CVE-2021-30341 & CVE-2022-25831 & CVE-2021-33626 & CVE-2022-23926 \\
CVE-2021-0187 & CVE-2022-25682 & CVE-2023-20589 & CVE-2021-46764 & CVE-2022-23927 \\
CVE-2022-26006 & CVE-2022-25695 & CVE-2023-21513 & CVE-2022-23820 & CVE-2022-23928 \\
CVE-2022-26837 & CVE-2022-25724 & CVE-2023-47616 & CVE-2022-23821 & CVE-2022-23930 \\
CVE-2022-32766 & CVE-2022-25727 & CVE-2023-52101 & CVE-2023-31342 & CVE-2022-23931 \\
CVE-2022-33894 & CVE-2023-26497 & CVE-2024-21483 & CVE-2023-31343 & CVE-2022-23932 \\
CVE-2023-22449 & CVE-2023-41111 & CVE-2024-53651 & CVE-2023-31345 & CVE-2022-23933 \\
CVE-2023-43758 & CVE-2023-41112 & CVE-2024-53832 & CVE-2022-39854 & CVE-2022-23934 \\
\hline
\end{tabular}%
}
\label{tab:centroid_cves}
\end{table}

After obtaining the hardware-related CVEs from 2021 to 2024, we began by embedding the 1{,}742 candidate CVE descriptions using the OpenAI model \texttt{text-embedding-ada-002}, resulting in contextualized semantic vectors suitable for downstream clustering. K-means clustering was then applied, with the number of clusters set to \(k=5\), as determined by the Elbow Method. To uncover the semantics within each cluster, we performed an n-gram frequency analysis over the CVE descriptions, extracting the top 15 most frequent uni-, bi-, and tri-grams. These keywords revealed prominent lexical patterns associated with distinct hardware vulnerability categories and were visualized as word clouds in Figure~\ref{fig:wordcloud_topics}, where font size indicates relative frequency. The keyword distributions exhibited clear distinctions across clusters.

Building on these statistical patterns, we remove noise such as brand and product names, and use the summarization procedure described in Section~\ref{sec:summarize} to generate concise, interpretable topic labels. For each cluster \(C_k\), the corresponding n-gram set \(\mathcal{W}_k\) was appended to the fixed prompt \(\rho_{\text{sum}}\) and input to the large language model \(\mathcal{M}\) (GPT-5). The model produced topic summaries \(T_k\) that abstracted root causes using accurate domain-relevant terminology. The five resulting topics are shown in Table~\ref{tab:topic_summaries}. We further visualized the clustering structure using t-distributed stochastic neighbor embedding (t-SNE) \cite{van2008visualizing}, which projected the high-dimensional vectors into two dimensions. As shown in Figure~\ref{fig:tsne_clusters}, each point corresponds to a CVE and is color-coded by its cluster assignment.

Lastly, to provide concrete examples of each cluster, we identified ten representative CVE entries closest to the cluster centroids based on cosine similarity in the embedding space. These entries are listed in Table~\ref{tab:centroid_cves}, which provides insight into the dominant patterns underlying each topic.


\section{Conclusion}
In this work, we presented LLM-HyPZ, a hybrid framework that leverages large language models and contextualized embeddings for zero-shot classification and clustering, enabling large-scale mining of cybersecurity corpora. Our experimental results demonstrate that the framework can effectively identify and organize hardware vulnerabilities into semantically coherent clusters without requiring labeled data, offering a scalable alternative to expert-driven methodologies.

Importantly, this framework directly supported the CWE Most Important Hardware Weaknesses (MIHW) 2025 update \cite{MITRE2025}. By applying LLM-HyPZ to the CVE corpus from 2021–2024, we systematically narrowed down the search space for human experts. Our method filtered and candidate CVEs, enabling reviewers to focus on a tractable subset.After human validation, 1,026 CVE records were selected as the basis for downstream analysis in the MIHW process, and our contribution accounts for 411 of these. In other words, our pipeline directly surfaced 411 of the 1,026 CVEs incorporated into the MIHW workflow, thereby reducing expert effort and accelerating evidence gathering. This demonstrates the value of AI-assisted data collection and analysis in establishing a more rigorous and scalable foundation for the MIHW selection process, complementing expert knowledge with data-driven evidence. The detailed list of identified HW CVE records is provided in Appendix \ref{app:discovered}.

Despite these strengths, several limitations remain. First, the contextual ambiguity between hardware, firmware, and software vulnerabilities is nontrivial, particularly in CVE entries that span multiple abstraction layers. While our binary classification achieved high accuracy, borderline cases still require human validation to ensure reliability. Second, the reliance on high-capacity models such as LLaMA 3.3 70B and GPT-4o introduces significant computational and financial overhead, which may limit the practicality of large-scale deployment. Finally, our clustering approach assumes a fixed number of clusters, which may not fully capture the evolving taxonomy of hardware weaknesses.

Looking forward, we aim to refine this approach by incorporating ensemble-based classification mechanisms (e.g., majority voting among multiple LLMs), multi-label and hierarchical categorization to capture hybrid vulnerabilities, and document-level zero-shot classification to improve granularity. We also envision lightweight inference pipelines that balance performance with efficiency, ensuring the scalability of LLM-assisted workflows in vulnerability mining. Together, these advancements will further strengthen the integration of AI-driven methods with community-driven standards such as the CWE, moving toward a sustainable and interpretable ecosystem for hardware security analysis.

\section*{Acknowledgment}
This work was partially supported by the National Science Foundation (NSF) under research project 2335046, the AI-HDL competition hosted by the University of Arizona, and the OpenAI Researcher Access Program 0000011862. We would like to thank Intel researchers Jason Fung, Arun Kanuparthi, and Hareesh Khattri for their valuable expert insights, which greatly informed the development of this work. The authors also acknowledge the MITRE CWE MIHW 2025 working group for their support and collaboration. Finally, we gratefully recognize the AI Verde team at the University of Arizona Data Science Institute for their support and resources, which enabled the use of LLaMA models in this study.

\bibliographystyle{IEEEtran} 
\bibliography{refs}

\begin{thebibliography}{10}
\providecommand{\url}[1]{#1}
\csname url@samestyle\endcsname
\providecommand{\newblock}{\relax}
\providecommand{\bibinfo}[2]{#2}
\providecommand{\BIBentrySTDinterwordspacing}{\spaceskip=0pt\relax}
\providecommand{\BIBentryALTinterwordstretchfactor}{4}
\providecommand{\BIBentryALTinterwordspacing}{\spaceskip=\fontdimen2\font plus
\BIBentryALTinterwordstretchfactor\fontdimen3\font minus \fontdimen4\font\relax}
\providecommand{\BIBforeignlanguage}[2]{{%
\expandafter\ifx\csname l@#1\endcsname\relax
\typeout{** WARNING: IEEEtran.bst: No hyphenation pattern has been}%
\typeout{** loaded for the language `#1'. Using the pattern for}%
\typeout{** the default language instead.}%
\else
\language=\csname l@#1\endcsname
\fi
#2}}
\providecommand{\BIBdecl}{\relax}
\BIBdecl

\bibitem{potlapally2011hardware}
N.~Potlapally, ``Hardware security in practice: Challenges and opportunities,'' in \emph{2011 IEEE International Symposium on Hardware-Oriented Security and Trust}.\hskip 1em plus 0.5em minus 0.4em\relax IEEE, 2011, pp. 93--98.

\bibitem{rosenfeld2010attacks}
K.~Rosenfeld and R.~Karri, ``Attacks and defenses for jtag,'' \emph{IEEE Design \& Test of Computers}, vol.~27, no.~1, pp. 36--47, 2010.

\bibitem{bojanova2024comprehensively}
I.~Bojanova, ``Comprehensively labeled weakness and vulnerability datasets via unambiguous formal bugs framework specifications,'' \emph{IT Professional}, vol.~26, no.~1, pp. 60--68, 2024.

\bibitem{MITRE2021}
\BIBentryALTinterwordspacing
MITRE, ``2021 cwe most important hardware weaknesses,'' Aug 2021. [Online]. Available: \url{https://cwe.mitre.org/topHW/archive/2021/2021_CWE_MIHW.html}
\BIBentrySTDinterwordspacing

\bibitem{dessouky2019hardfails}
G.~Dessouky, D.~Gens, P.~Haney, G.~Persyn, A.~Kanuparthi, H.~Khattri, J.~M. Fung, A.-R. Sadeghi, and J.~Rajendran, ``$\{$HardFails$\}$: insights into $\{$software-exploitable$\}$ hardware bugs,'' in \emph{28th USENIX Security Symposium (USENIX Security 19)}, 2019, pp. 213--230.

\bibitem{chen2018categorizing}
Q.~Chen, L.~Bao, L.~Li, X.~Xia, and L.~Cai, ``Categorizing and predicting invalid vulnerabilities on common vulnerabilities and exposures,'' in \emph{2018 25th Asia-Pacific Software Engineering Conference (APSEC)}.\hskip 1em plus 0.5em minus 0.4em\relax IEEE, 2018, pp. 345--354.

\bibitem{waareus2020automated}
E.~W{\aa}reus and M.~Hell, ``Automated cpe labeling of cve summaries with machine learning,'' in \emph{Detection of Intrusions and Malware, and Vulnerability Assessment: 17th International Conference, DIMVA 2020, Lisbon, Portugal, June 24--26, 2020, Proceedings 17}.\hskip 1em plus 0.5em minus 0.4em\relax Springer, 2020, pp. 3--22.

\bibitem{yitagesu2021unsupervised}
S.~Yitagesu, Z.~Xing, X.~Zhang, Z.~Feng, X.~Li, and L.~Han, ``Unsupervised labeling and extraction of phrase-based concepts in vulnerability descriptions,'' in \emph{2021 36th IEEE/ACM International Conference on Automated Software Engineering (ASE)}.\hskip 1em plus 0.5em minus 0.4em\relax IEEE, 2021, pp. 943--954.

\bibitem{chen2002data}
M.-S. Chen, J.~Han, and P.~S. Yu, ``Data mining: an overview from a database perspective,'' \emph{IEEE Transactions on Knowledge and data Engineering}, vol.~8, no.~6, pp. 866--883, 2002.

\bibitem{kojima2022large}
T.~Kojima, S.~S. Gu, M.~Reid, Y.~Matsuo, and Y.~Iwasawa, ``Large language models are zero-shot reasoners,'' \emph{Advances in neural information processing systems}, vol.~35, pp. 22\,199--22\,213, 2022.

\bibitem{fink2023potential}
M.~A. Fink, A.~Bischoff, C.~A. Fink, M.~Moll, J.~Kroschke, L.~Dulz, C.~P. Heu{\ss}el, H.-U. Kauczor, and T.~F. Weber, ``Potential of chatgpt and gpt-4 for data mining of free-text ct reports on lung cancer,'' \emph{Radiology}, vol. 308, no.~3, p. e231362, 2023.

\bibitem{tang2023does}
R.~Tang, X.~Han, X.~Jiang, and X.~Hu, ``Does synthetic data generation of llms help clinical text mining?'' \emph{arXiv preprint arXiv:2303.04360}, 2023.

\bibitem{khoury2021analysis}
R.~Khoury, B.~Vignau, S.~Hall{\'e}, A.~Hamou-Lhadj, and A.~Razgallah, ``An analysis of the use of cves by iot malware,'' in \emph{Foundations and Practice of Security: 13th International Symposium, FPS 2020, Montreal, QC, Canada, December 1--3, 2020, Revised Selected Papers 13}.\hskip 1em plus 0.5em minus 0.4em\relax Springer, 2021, pp. 47--62.

\bibitem{satam2020wids}
P.~Satam and S.~Hariri, ``Wids: An anomaly based intrusion detection system for wi-fi (ieee 802.11) protocol,'' \emph{IEEE Transactions on Network and Service Management}, vol.~18, no.~1, pp. 1077--1091, 2020.

\bibitem{satam2018bluetooth}
P.~Satam, S.~Satam, and S.~Hariri, ``Bluetooth intrusion detection system (bids),'' in \emph{2018 IEEE/ACS 15th International Conference on Computer Systems and Applications (AICCSA)}.\hskip 1em plus 0.5em minus 0.4em\relax IEEE, 2018, pp. 1--7.

\bibitem{satam2015anomaly}
P.~Satam, H.~R. Alipour, Y.~B. Al-Nashif, and S.~Hariri, ``Anomaly behavior analysis of dns protocol.'' \emph{J. Internet Serv. Inf. Secur.}, vol.~5, no.~4, pp. 85--97, 2015.

\bibitem{bhurtel2023unveiling}
M.~Bhurtel and D.~B. Rawat, ``Unveiling the landscape of operating system vulnerabilities,'' \emph{Future Internet}, vol.~15, no.~7, p. 248, 2023.

\bibitem{neuhaus2010security}
S.~Neuhaus and T.~Zimmermann, ``Security trend analysis with cve topic models,'' in \emph{2010 IEEE 21st International Symposium on Software Reliability Engineering}.\hskip 1em plus 0.5em minus 0.4em\relax IEEE, 2010, pp. 111--120.

\bibitem{li2020survey}
J.~Li, A.~Sun, J.~Han, and C.~Li, ``A survey on deep learning for named entity recognition,'' \emph{IEEE transactions on knowledge and data engineering}, vol.~34, no.~1, pp. 50--70, 2020.

\bibitem{sutton2012introduction}
C.~Sutton, A.~McCallum \emph{et~al.}, ``An introduction to conditional random fields,'' \emph{Foundations and Trends{\textregistered} in Machine Learning}, vol.~4, no.~4, pp. 267--373, 2012.

\bibitem{zhang2025teleclass}
Y.~Zhang, R.~Yang, X.~Xu, R.~Li, J.~Xiao, J.~Shen, and J.~Han, ``Teleclass: Taxonomy enrichment and llm-enhanced hierarchical text classification with minimal supervision,'' in \emph{Proceedings of the ACM on Web Conference 2025}, 2025, pp. 2032--2042.

\bibitem{wan2024tnt}
M.~Wan, T.~Safavi, S.~K. Jauhar, Y.~Kim, S.~Counts, J.~Neville, S.~Suri, C.~Shah, R.~W. White, L.~Yang \emph{et~al.}, ``Tnt-llm: Text mining at scale with large language models,'' in \emph{Proceedings of the 30th ACM SIGKDD conference on knowledge discovery and data mining}, 2024, pp. 5836--5847.

\bibitem{ma2025should}
Q.~Ma, W.~Peng, C.~Yang, H.~Shen, K.~Koedinger, and T.~Wu, ``What should we engineer in prompts? training humans in requirement-driven llm use,'' \emph{ACM Transactions on Computer-Human Interaction}, 2025.

\bibitem{cai2013multi}
X.~Cai, F.~Nie, and H.~Huang, ``Multi-view k-means clustering on big data.'' in \emph{IJCAI}, vol.~13, 2013, pp. 2598--2604.

\bibitem{white2023prompt}
J.~White, Q.~Fu, S.~Hays, M.~Sandborn, C.~Olea, H.~Gilbert, A.~Elnashar, J.~Spencer-Smith, and D.~C. Schmidt, ``A prompt pattern catalog to enhance prompt engineering with chatgpt,'' \emph{arXiv preprint arXiv:2302.11382}, 2023.

\bibitem{OpenAI}
\BIBentryALTinterwordspacing
OpenAI, ``Openai python sdk.'' [Online]. Available: \url{https://platform.openai.com/docs/api-reference/introduction?lang=python}
\BIBentrySTDinterwordspacing

\bibitem{mithun2025ai}
P.~Mithun, E.~Noriega-Atala, N.~Merchant, and E.~Skidmore, ``Ai-verde: A gateway for egalitarian access to large language model-based resources for educational institutions,'' \emph{arXiv preprint arXiv:2502.09651}, 2025.

\bibitem{van2008visualizing}
L.~Van~der Maaten and G.~Hinton, ``Visualizing data using t-sne.'' \emph{Journal of machine learning research}, vol.~9, no.~11, 2008.

\bibitem{MITRE2025}
\BIBentryALTinterwordspacing
MITRE, ``2025 cwe most important hardware weaknesses,'' Aug 2025. [Online]. Available: \url{https://cwe.mitre.org/topHW/archive/2025/2025_CWE_MIHW.html}
\BIBentrySTDinterwordspacing

\end{thebibliography}

\appendices
\section{Validation Dataset ($\mathcal{D}_{\text{CVE\_Validation}}$) for Zero-shot HW/SW Classfication Task} \label{app:validation_dataset}
The hardware and software CVE entries for validation are shown in Table \ref{tab:validation}.

\begin{table*}[]
\caption{Validation Set for Classification Task}
\resizebox{\textwidth}{!}{%
\begin{tabular}{|lllll|lllll|}
\hline
\multicolumn{5}{|c|}{Hardware CVE Entries} &
  \multicolumn{5}{c|}{Software CVE Entries} \\ \hline
\multicolumn{1}{|l|}{CVE-2020-12954} &
  \multicolumn{1}{l|}{CVE-2020-12961} &
  \multicolumn{1}{l|}{CVE-2020-12965} &
  \multicolumn{1}{l|}{CVE-2020-12966} &
  CVE-2020-12967 &
  \multicolumn{1}{l|}{CVE-2024-0001} &
  \multicolumn{1}{l|}{CVE-2024-0002} &
  \multicolumn{1}{l|}{CVE-2024-0003} &
  \multicolumn{1}{l|}{CVE-2024-0004} &
  CVE-2024-0005 \\ \hline
\multicolumn{1}{|l|}{CVE-2020-12988} &
  \multicolumn{1}{l|}{CVE-2020-24489} &
  \multicolumn{1}{l|}{CVE-2020-24511} &
  \multicolumn{1}{l|}{CVE-2020-24513} &
  CVE-2020-24516 &
  \multicolumn{1}{l|}{CVE-2024-0006} &
  \multicolumn{1}{l|}{CVE-2024-0007} &
  \multicolumn{1}{l|}{CVE-2024-0008} &
  \multicolumn{1}{l|}{CVE-2024-0009} &
  CVE-2024-0010 \\ \hline
\multicolumn{1}{|l|}{CVE-2021-0144} &
  \multicolumn{1}{l|}{CVE-2021-0145} &
  \multicolumn{1}{l|}{CVE-2021-0146} &
  \multicolumn{1}{l|}{CVE‑2021‑1088} &
  CVE‑2021‑1105 &
  \multicolumn{1}{l|}{CVE-2024-0011} &
  \multicolumn{1}{l|}{CVE-2024-0012} &
  \multicolumn{1}{l|}{CVE-2024-0014} &
  \multicolumn{1}{l|}{CVE-2024-0015} &
  CVE-2024-0016 \\ \hline
\multicolumn{1}{|l|}{CVE‑2021‑1125} &
  \multicolumn{1}{l|}{CVE‑2021‑23201} &
  \multicolumn{1}{l|}{CVE‑2021‑23217} &
  \multicolumn{1}{l|}{CVE‑2021‑23219} &
  CVE-2021-26311 &
  \multicolumn{1}{l|}{CVE-2024-0017} &
  \multicolumn{1}{l|}{CVE-2024-0018} &
  \multicolumn{1}{l|}{CVE-2024-0019} &
  \multicolumn{1}{l|}{CVE-2024-0020} &
  CVE-2024-0021 \\ \hline
\multicolumn{1}{|l|}{CVE-2021-26312} &
  \multicolumn{1}{l|}{CVE-2021-26312} &
  \multicolumn{1}{l|}{CVE-2021-26313} &
  \multicolumn{1}{l|}{CVE-2021-26314} &
  CVE-2021-26318 &
  \multicolumn{1}{l|}{CVE-2024-0022} &
  \multicolumn{1}{l|}{CVE-2024-0023} &
  \multicolumn{1}{l|}{CVE-2024-0024} &
  \multicolumn{1}{l|}{CVE-2024-0025} &
  CVE-2024-0026 \\ \hline
\multicolumn{1}{|l|}{CVE-2021-26324} &
  \multicolumn{1}{l|}{CVE-2021-26331} &
  \multicolumn{1}{l|}{CVE-2021-26332} &
  \multicolumn{1}{l|}{CVE-2021-26336} &
  CVE-2021-26337 &
  \multicolumn{1}{l|}{CVE-2024-0027} &
  \multicolumn{1}{l|}{CVE-2024-0029} &
  \multicolumn{1}{l|}{CVE-2024-0030} &
  \multicolumn{1}{l|}{CVE-2024-0031} &
  CVE-2024-0032 \\ \hline
\multicolumn{1}{|l|}{CVE-2021-26337} &
  \multicolumn{1}{l|}{CVE-2021-26338} &
  \multicolumn{1}{l|}{CVE-2021-26339} &
  \multicolumn{1}{l|}{CVE-2021-26340} &
  CVE-2021-26341 &
  \multicolumn{1}{l|}{CVE-2024-0033} &
  \multicolumn{1}{l|}{CVE-2024-0034} &
  \multicolumn{1}{l|}{CVE-2024-0035} &
  \multicolumn{1}{l|}{CVE-2024-0036} &
  CVE-2024-0037 \\ \hline
\multicolumn{1}{|l|}{CVE-2021-26342} &
  \multicolumn{1}{l|}{CVE-2021-26345} &
  \multicolumn{1}{l|}{CVE-2021-26348} &
  \multicolumn{1}{l|}{CVE-2021-26349} &
  CVE-2021-26351 &
  \multicolumn{1}{l|}{CVE-2024-0038} &
  \multicolumn{1}{l|}{CVE-2024-0039} &
  \multicolumn{1}{l|}{CVE-2024-0040} &
  \multicolumn{1}{l|}{CVE-2024-0041} &
  CVE-2024-0042 \\ \hline
\multicolumn{1}{|l|}{CVE-2021-26351} &
  \multicolumn{1}{l|}{CVE-2021-26352} &
  \multicolumn{1}{l|}{CVE-2021-26355} &
  \multicolumn{1}{l|}{CVE-2021-26366} &
  CVE-2021-26367 &
  \multicolumn{1}{l|}{CVE-2024-0043} &
  \multicolumn{1}{l|}{CVE-2024-0044} &
  \multicolumn{1}{l|}{CVE-2024-0045} &
  \multicolumn{1}{l|}{CVE-2024-0046} &
  CVE-2024-0047 \\ \hline
\multicolumn{1}{|l|}{CVE-2021-26373} &
  \multicolumn{1}{l|}{CVE-2021-26375} &
  \multicolumn{1}{l|}{CVE-2021-26393} &
  \multicolumn{1}{l|}{CVE-2021-26393} &
  CVE-2021-26396 &
  \multicolumn{1}{l|}{CVE-2024-0048} &
  \multicolumn{1}{l|}{CVE-2024-0049} &
  \multicolumn{1}{l|}{CVE-2024-0050} &
  \multicolumn{1}{l|}{CVE-2024-0051} &
  CVE-2024-0052 \\ \hline
\multicolumn{1}{|l|}{CVE-2021-26400} &
  \multicolumn{1}{l|}{CVE-2021-26407} &
  \multicolumn{1}{l|}{CVE-2021-33096} &
  \multicolumn{1}{l|}{CVE-2021-33149} &
  CVE-2021-33150 &
  \multicolumn{1}{l|}{CVE-2024-0053} &
  \multicolumn{1}{l|}{CVE-2024-0054} &
  \multicolumn{1}{l|}{CVE-2024-0055} &
  \multicolumn{1}{l|}{CVE-2024-0056} &
  CVE-2024-0057 \\ \hline
\multicolumn{1}{|l|}{CVE‑2021‑34399} &
  \multicolumn{1}{l|}{CVE‑2021‑34400} &
  \multicolumn{1}{l|}{CVE-2021-46744} &
  \multicolumn{1}{l|}{CVE-2021-46746} &
  CVE-2021-46764 &
  \multicolumn{1}{l|}{CVE-2024-0066} &
  \multicolumn{1}{l|}{CVE-2024-0067} &
  \multicolumn{1}{l|}{CVE-2024-0068} &
  \multicolumn{1}{l|}{CVE-2024-0071} &
  CVE-2024-0072 \\ \hline
\multicolumn{1}{|l|}{CVE-2021-46778} &
  \multicolumn{1}{l|}{CVE-2022-0001} &
  \multicolumn{1}{l|}{CVE-2022-0002} &
  \multicolumn{1}{l|}{CVE-2022-0005} &
  CVE-2022-21123 &
  \multicolumn{1}{l|}{CVE-2024-0073} &
  \multicolumn{1}{l|}{CVE-2024-0074} &
  \multicolumn{1}{l|}{CVE-2024-0075} &
  \multicolumn{1}{l|}{CVE-2024-0076} &
  CVE-2024-0077 \\ \hline
\multicolumn{1}{|l|}{CVE-2022-21125} &
  \multicolumn{1}{l|}{CVE-2022-21127} &
  \multicolumn{1}{l|}{CVE-2022-21131} &
  \multicolumn{1}{l|}{CVE-2022-21151} &
  CVE-2022-21166 &
  \multicolumn{1}{l|}{CVE-2024-0078} &
  \multicolumn{1}{l|}{CVE-2024-0079} &
  \multicolumn{1}{l|}{CVE-2024-0080} &
  \multicolumn{1}{l|}{CVE-2024-0081} &
  CVE-2024-0082 \\ \hline
\multicolumn{1}{|l|}{CVE-2022-21180} &
  \multicolumn{1}{l|}{CVE-2022-21216} &
  \multicolumn{1}{l|}{CVE-2022-21233} &
  \multicolumn{1}{l|}{CVE-2022-23818} &
  CVE-2022-23821 &
  \multicolumn{1}{l|}{CVE-2024-0083} &
  \multicolumn{1}{l|}{CVE-2024-0084} &
  \multicolumn{1}{l|}{CVE-2024-0085} &
  \multicolumn{1}{l|}{CVE-2024-0086} &
  CVE-2024-0087 \\ \hline
\multicolumn{1}{|l|}{CVE-2022-23823} &
  \multicolumn{1}{l|}{CVE-2022-23824} &
  \multicolumn{1}{l|}{CVE-2022-23825} &
  \multicolumn{1}{l|}{CVE-2022-23829} &
  CVE-2022-23830 &
  \multicolumn{1}{l|}{CVE-2024-0088} &
  \multicolumn{1}{l|}{CVE-2024-0089} &
  \multicolumn{1}{l|}{CVE-2024-0090} &
  \multicolumn{1}{l|}{CVE-2024-0091} &
  CVE-2024-0092 \\ \hline
\multicolumn{1}{|l|}{CVE-2022-26373} &
  \multicolumn{1}{l|}{CVE-2022-27672} &
  \multicolumn{1}{l|}{CVE-2022-28693} &
  \multicolumn{1}{l|}{CVE-2022-29900} &
  CVE-2022-29901 &
  \multicolumn{1}{l|}{CVE-2024-0093} &
  \multicolumn{1}{l|}{CVE-2024-0094} &
  \multicolumn{1}{l|}{CVE-2024-0095} &
  \multicolumn{1}{l|}{CVE-2024-0096} &
  CVE-2024-0097 \\ \hline
\multicolumn{1}{|l|}{CVE-2022-33196} &
  \multicolumn{1}{l|}{CVE-2022-33972} &
  \multicolumn{1}{l|}{CVE-2022-38090} &
  \multicolumn{1}{l|}{CVE-2022-40982} &
  CVE-2022-41804 &
  \multicolumn{1}{l|}{CVE-2024-0098} &
  \multicolumn{1}{l|}{CVE-2024-0099} &
  \multicolumn{1}{l|}{CVE-2024-0100} &
  \multicolumn{1}{l|}{CVE-2024-0101} &
  CVE-2024-0102 \\ \hline
\multicolumn{1}{|l|}{CVE-2023-20518} &
  \multicolumn{1}{l|}{CVE-2023-20524} &
  \multicolumn{1}{l|}{CVE-2023-20533} &
  \multicolumn{1}{l|}{CVE-2023-20566} &
  CVE-2023-20569 &
  \multicolumn{1}{l|}{CVE-2024-0103} &
  \multicolumn{1}{l|}{CVE-2024-0104} &
  \multicolumn{1}{l|}{CVE-2024-0105} &
  \multicolumn{1}{l|}{CVE-2024-0106} &
  CVE-2024-0107 \\ \hline
\multicolumn{1}{|l|}{CVE-2023-20570} &
  \multicolumn{1}{l|}{CVE-2023-20573} &
  \multicolumn{1}{l|}{CVE-2023-20575} &
  \multicolumn{1}{l|}{CVE-2023-20579} &
  CVE-2023-20583 &
  \multicolumn{1}{l|}{CVE-2024-0108} &
  \multicolumn{1}{l|}{CVE-2024-0109} &
  \multicolumn{1}{l|}{CVE-2024-0110} &
  \multicolumn{1}{l|}{CVE-2024-0111} &
  CVE-2024-0112 \\ \hline
\end{tabular}%
}
\label{tab:validation}
\end{table*}

\section{} \label{app:discovered}
The Table \ref{tab:MIHW_List} presents the subset of CVE records identified by the proposed LLM-HyPZ framework as hardware-related vulnerabilities. These records were subsequently incorporated into the foundation of the MITRE CWE Most Important Hardware Weaknesses (MIHW) 2025 analysis.

\begin{table*}[]
\caption{CVE Records Identified by LLM-HyPZ in Support of the MITRE MIHW 2025}
\resizebox{\textwidth}{!}{%
\begin{tabular}{|llllllllll|}
\hline
\multicolumn{10}{|c|}{HW Issues} \\ \hline
\multicolumn{1}{|l|}{CVE-2021-1925} &
  \multicolumn{1}{l|}{CVE-2021-1938} &
  \multicolumn{1}{l|}{CVE-2021-1971} &
  \multicolumn{1}{l|}{CVE-2021-20168} &
  \multicolumn{1}{l|}{CVE-2021-20826} &
  \multicolumn{1}{l|}{CVE-2021-20870} &
  \multicolumn{1}{l|}{CVE-2021-27477} &
  \multicolumn{1}{l|}{CVE-2021-30328} &
  \multicolumn{1}{l|}{CVE-2021-30336} &
  CVE-2021-30350 \\ \hline
\multicolumn{1}{|l|}{CVE-2021-31532} &
  \multicolumn{1}{l|}{CVE-2021-31612} &
  \multicolumn{1}{l|}{CVE-2021-33149} &
  \multicolumn{1}{l|}{CVE-2021-33887} &
  \multicolumn{1}{l|}{CVE-2021-35465} &
  \multicolumn{1}{l|}{CVE-2021-36315} &
  \multicolumn{1}{l|}{CVE-2021-37001} &
  \multicolumn{1}{l|}{CVE-2021-37273} &
  \multicolumn{1}{l|}{CVE-2021-3789} &
  CVE-2021-38365 \\ \hline
\multicolumn{1}{|l|}{CVE-2021-38392} &
  \multicolumn{1}{l|}{CVE-2021-38400} &
  \multicolumn{1}{l|}{CVE-2021-38543} &
  \multicolumn{1}{l|}{CVE-2021-38548} &
  \multicolumn{1}{l|}{CVE-2021-38549} &
  \multicolumn{1}{l|}{CVE-2021-40368} &
  \multicolumn{1}{l|}{CVE-2021-42114} &
  \multicolumn{1}{l|}{CVE-2021-46744} &
  \multicolumn{1}{l|}{CVE-2021-46774} &
  CVE-2021-46778 \\ \hline
\multicolumn{1}{|l|}{CVE-2022-0001} &
  \multicolumn{1}{l|}{CVE-2022-0002} &
  \multicolumn{1}{l|}{CVE-2022-1716} &
  \multicolumn{1}{l|}{CVE-2022-22083} &
  \multicolumn{1}{l|}{CVE-2022-23823} &
  \multicolumn{1}{l|}{CVE-2022-23825} &
  \multicolumn{1}{l|}{CVE-2022-26390} &
  \multicolumn{1}{l|}{CVE-2022-27250} &
  \multicolumn{1}{l|}{CVE-2022-2741} &
  CVE-2022-28383 \\ \hline
\multicolumn{1}{|l|}{CVE-2022-28386} &
  \multicolumn{1}{l|}{CVE-2022-30467} &
  \multicolumn{1}{l|}{CVE-2022-31756} &
  \multicolumn{1}{l|}{CVE-2022-33972} &
  \multicolumn{1}{l|}{CVE-2022-34633} &
  \multicolumn{1}{l|}{CVE-2022-37939} &
  \multicolumn{1}{l|}{CVE-2022-38090} &
  \multicolumn{1}{l|}{CVE-2022-38392} &
  \multicolumn{1}{l|}{CVE-2022-38465} &
  CVE-2022-38766 \\ \hline
\multicolumn{1}{|l|}{CVE-2022-38788} &
  \multicolumn{1}{l|}{CVE-2022-38973} &
  \multicolumn{1}{l|}{CVE-2022-39064} &
  \multicolumn{1}{l|}{CVE-2022-39065} &
  \multicolumn{1}{l|}{CVE-2022-39854} &
  \multicolumn{1}{l|}{CVE-2022-39901} &
  \multicolumn{1}{l|}{CVE-2022-41984} &
  \multicolumn{1}{l|}{CVE-2022-45552} &
  \multicolumn{1}{l|}{CVE-2022-45914} &
  CVE-2022-46142 \\ \hline
\multicolumn{1}{|l|}{CVE-2022-46740} &
  \multicolumn{1}{l|}{CVE-2022-47036} &
  \multicolumn{1}{l|}{CVE-2022-47976} &
  \multicolumn{1}{l|}{CVE-2023-20584} &
  \multicolumn{1}{l|}{CVE-2023-20587} &
  \multicolumn{1}{l|}{CVE-2023-20588} &
  \multicolumn{1}{l|}{CVE-2023-20589} &
  \multicolumn{1}{l|}{CVE-2023-20591} &
  \multicolumn{1}{l|}{CVE-2023-20593} &
  CVE-2023-21513 \\ \hline
\multicolumn{1}{|l|}{CVE-2023-23774} &
  \multicolumn{1}{l|}{CVE-2023-26282} &
  \multicolumn{1}{l|}{CVE-2023-26497} &
  \multicolumn{1}{l|}{CVE-2023-26498} &
  \multicolumn{1}{l|}{CVE-2023-26943} &
  \multicolumn{1}{l|}{CVE-2023-28092} &
  \multicolumn{1}{l|}{CVE-2023-29086} &
  \multicolumn{1}{l|}{CVE-2023-29087} &
  \multicolumn{1}{l|}{CVE-2023-29088} &
  CVE-2023-29089 \\ \hline
\multicolumn{1}{|l|}{CVE-2023-29090} &
  \multicolumn{1}{l|}{CVE-2023-29091} &
  \multicolumn{1}{l|}{CVE-2023-29092} &
  \multicolumn{1}{l|}{CVE-2023-29389} &
  \multicolumn{1}{l|}{CVE-2023-30354} &
  \multicolumn{1}{l|}{CVE-2023-30561} &
  \multicolumn{1}{l|}{CVE-2023-32219} &
  \multicolumn{1}{l|}{CVE-2023-33921} &
  \multicolumn{1}{l|}{CVE-2023-34404} &
  CVE-2023-35818 \\ \hline
\multicolumn{1}{|l|}{CVE-2023-37377} &
  \multicolumn{1}{l|}{CVE-2023-39368} &
  \multicolumn{1}{l|}{CVE-2023-39841} &
  \multicolumn{1}{l|}{CVE-2023-39842} &
  \multicolumn{1}{l|}{CVE-2023-43757} &
  \multicolumn{1}{l|}{CVE-2023-50807} &
  \multicolumn{1}{l|}{CVE-2023-52043} &
  \multicolumn{1}{l|}{CVE-2023-5449} &
  \multicolumn{1}{l|}{CVE-2024-21460} &
  CVE-2024-21820 \\ \hline
\multicolumn{1}{|l|}{CVE-2024-2193} &
  \multicolumn{1}{l|}{CVE-2024-22006} &
  \multicolumn{1}{l|}{CVE-2024-22044} &
  \multicolumn{1}{l|}{CVE-2024-25077} &
  \multicolumn{1}{l|}{CVE-2024-25883} &
  \multicolumn{1}{l|}{CVE-2024-25972} &
  \multicolumn{1}{l|}{CVE-2024-26790} &
  \multicolumn{1}{l|}{CVE-2024-27360} &
  \multicolumn{1}{l|}{CVE-2024-28067} &
  CVE-2024-29153 \\ \hline
\multicolumn{1}{|l|}{CVE-2024-30189} &
  \multicolumn{1}{l|}{CVE-2024-31142} &
  \multicolumn{1}{l|}{CVE-2024-31799} &
  \multicolumn{1}{l|}{CVE-2024-31800} &
  \multicolumn{1}{l|}{CVE-2024-31957} &
  \multicolumn{1}{l|}{CVE-2024-31959} &
  \multicolumn{1}{l|}{CVE-2024-35311} &
  \multicolumn{1}{l|}{CVE-2024-36242} &
  \multicolumn{1}{l|}{CVE-2024-36293} &
  CVE-2024-37649 \\ \hline
\multicolumn{1}{|l|}{CVE-2024-38280} &
  \multicolumn{1}{l|}{CVE-2024-38283} &
  \multicolumn{1}{l|}{CVE-2024-38404} &
  \multicolumn{1}{l|}{CVE-2024-38660} &
  \multicolumn{1}{l|}{CVE-2024-39278} &
  \multicolumn{1}{l|}{CVE-2024-39922} &
  \multicolumn{1}{l|}{CVE-2024-41927} &
  \multicolumn{1}{l|}{CVE-2024-42095} &
  \multicolumn{1}{l|}{CVE-2024-44067} &
  CVE-2024-44815 \\ \hline
\multicolumn{1}{|l|}{CVE-2024-45368} &
  \multicolumn{1}{l|}{CVE-2024-49414} &
  \multicolumn{1}{l|}{CVE-2024-49422} &
  \multicolumn{1}{l|}{CVE-2024-49834} &
  \multicolumn{1}{l|}{CVE-2024-51072} &
  \multicolumn{1}{l|}{CVE-2024-51073} &
  \multicolumn{1}{l|}{CVE-2024-51074} &
  \multicolumn{1}{l|}{CVE-2024-53114} &
  \multicolumn{1}{l|}{CVE-2024-53651} &
   \\ \hline
\multicolumn{10}{|c|}{HW Devices} \\ \hline
\multicolumn{1}{|l|}{CVE-2021-0086} &
  \multicolumn{1}{l|}{CVE-2021-0089} &
  \multicolumn{1}{l|}{CVE-2021-0146} &
  \multicolumn{1}{l|}{CVE-2021-1104} &
  \multicolumn{1}{l|}{CVE-2021-1441} &
  \multicolumn{1}{l|}{CVE-2021-1888} &
  \multicolumn{1}{l|}{CVE-2021-1918} &
  \multicolumn{1}{l|}{CVE-2021-1924} &
  \multicolumn{1}{l|}{CVE-2021-1929} &
  CVE-2021-1957 \\ \hline
\multicolumn{1}{|l|}{CVE-2021-1960} &
  \multicolumn{1}{l|}{CVE-2021-20107} &
  \multicolumn{1}{l|}{CVE-2021-20161} &
  \multicolumn{1}{l|}{CVE-2021-20872} &
  \multicolumn{1}{l|}{CVE-2021-21739} &
  \multicolumn{1}{l|}{CVE-2021-22313} &
  \multicolumn{1}{l|}{CVE-2021-22317} &
  \multicolumn{1}{l|}{CVE-2021-22372} &
  \multicolumn{1}{l|}{CVE-2021-22373} &
  CVE-2021-22436 \\ \hline
\multicolumn{1}{|l|}{CVE-2021-22443} &
  \multicolumn{1}{l|}{CVE-2021-22744} &
  \multicolumn{1}{l|}{CVE-2021-23147} &
  \multicolumn{1}{l|}{CVE-2021-26312} &
  \multicolumn{1}{l|}{CVE-2021-26313} &
  \multicolumn{1}{l|}{CVE-2021-26314} &
  \multicolumn{1}{l|}{CVE-2021-26318} &
  \multicolumn{1}{l|}{CVE-2021-26324} &
  \multicolumn{1}{l|}{CVE-2021-26326} &
  CVE-2021-26328 \\ \hline
\multicolumn{1}{|l|}{CVE-2021-26342} &
  \multicolumn{1}{l|}{CVE-2021-26348} &
  \multicolumn{1}{l|}{CVE-2021-26360} &
  \multicolumn{1}{l|}{CVE-2021-26366} &
  \multicolumn{1}{l|}{CVE-2021-26367} &
  \multicolumn{1}{l|}{CVE-2021-26393} &
  \multicolumn{1}{l|}{CVE-2021-26400} &
  \multicolumn{1}{l|}{CVE-2021-26401} &
  \multicolumn{1}{l|}{CVE-2021-27458} &
  CVE-2021-27952 \\ \hline
\multicolumn{1}{|l|}{CVE-2021-29414} &
  \multicolumn{1}{l|}{CVE-2021-30061} &
  \multicolumn{1}{l|}{CVE-2021-30281} &
  \multicolumn{1}{l|}{CVE-2021-30302} &
  \multicolumn{1}{l|}{CVE-2021-30307} &
  \multicolumn{1}{l|}{CVE-2021-30322} &
  \multicolumn{1}{l|}{CVE-2021-30325} &
  \multicolumn{1}{l|}{CVE-2021-31345} &
  \multicolumn{1}{l|}{CVE-2021-31613} &
  CVE-2021-31785 \\ \hline
\multicolumn{1}{|l|}{CVE-2021-31787} &
  \multicolumn{1}{l|}{CVE-2021-33012} &
  \multicolumn{1}{l|}{CVE-2021-33096} &
  \multicolumn{1}{l|}{CVE-2021-33881} &
  \multicolumn{1}{l|}{CVE-2021-34174} &
  \multicolumn{1}{l|}{CVE-2021-35954} &
  \multicolumn{1}{l|}{CVE-2021-3704} &
  \multicolumn{1}{l|}{CVE-2021-3705} &
  \multicolumn{1}{l|}{CVE-2021-37110} &
  CVE-2021-37436 \\ \hline
\multicolumn{1}{|l|}{CVE-2021-38544} &
  \multicolumn{1}{l|}{CVE-2021-38545} &
  \multicolumn{1}{l|}{CVE-2021-38546} &
  \multicolumn{1}{l|}{CVE-2021-38547} &
  \multicolumn{1}{l|}{CVE-2021-39237} &
  \multicolumn{1}{l|}{CVE-2021-39363} &
  \multicolumn{1}{l|}{CVE-2021-39364} &
  \multicolumn{1}{l|}{CVE-2021-39729} &
  \multicolumn{1}{l|}{CVE-2021-40170} &
  CVE-2021-40171 \\ \hline
\multicolumn{1}{|l|}{CVE-2021-40506} &
  \multicolumn{1}{l|}{CVE-2021-40507} &
  \multicolumn{1}{l|}{CVE-2021-41612} &
  \multicolumn{1}{l|}{CVE-2021-41613} &
  \multicolumn{1}{l|}{CVE-2021-41614} &
  \multicolumn{1}{l|}{CVE-2021-42849} &
  \multicolumn{1}{l|}{CVE-2021-43282} &
  \multicolumn{1}{l|}{CVE-2021-43327} &
  \multicolumn{1}{l|}{CVE-2021-43392} &
  CVE-2021-46145 \\ \hline
\multicolumn{1}{|l|}{CVE-2022-0004} &
  \multicolumn{1}{l|}{CVE-2022-0005} &
  \multicolumn{1}{l|}{CVE-2022-1318} &
  \multicolumn{1}{l|}{CVE-2022-1955} &
  \multicolumn{1}{l|}{CVE-2022-20152} &
  \multicolumn{1}{l|}{CVE-2022-20660} &
  \multicolumn{1}{l|}{CVE-2022-20817} &
  \multicolumn{1}{l|}{CVE-2022-21123} &
  \multicolumn{1}{l|}{CVE-2022-21125} &
  CVE-2022-21127 \\ \hline
\multicolumn{1}{|l|}{CVE-2022-21131} &
  \multicolumn{1}{l|}{CVE-2022-21151} &
  \multicolumn{1}{l|}{CVE-2022-21166} &
  \multicolumn{1}{l|}{CVE-2022-21233} &
  \multicolumn{1}{l|}{CVE-2022-22154} &
  \multicolumn{1}{l|}{CVE-2022-23818} &
  \multicolumn{1}{l|}{CVE-2022-23822} &
  \multicolumn{1}{l|}{CVE-2022-23824} &
  \multicolumn{1}{l|}{CVE-2022-23829} &
  CVE-2022-23960 \\ \hline
\multicolumn{1}{|l|}{CVE-2022-24436} &
  \multicolumn{1}{l|}{CVE-2022-25915} &
  \multicolumn{1}{l|}{CVE-2022-26269} &
  \multicolumn{1}{l|}{CVE-2022-26296} &
  \multicolumn{1}{l|}{CVE-2022-26373} &
  \multicolumn{1}{l|}{CVE-2022-2675} &
  \multicolumn{1}{l|}{CVE-2022-27178} &
  \multicolumn{1}{l|}{CVE-2022-27254} &
  \multicolumn{1}{l|}{CVE-2022-27660} &
  CVE-2022-27672 \\ \hline
\multicolumn{1}{|l|}{CVE-2022-27948} &
  \multicolumn{1}{l|}{CVE-2022-28384} &
  \multicolumn{1}{l|}{CVE-2022-28385} &
  \multicolumn{1}{l|}{CVE-2022-28387} &
  \multicolumn{1}{l|}{CVE-2022-28693} &
  \multicolumn{1}{l|}{CVE-2022-29402} &
  \multicolumn{1}{l|}{CVE-2022-29792} &
  \multicolumn{1}{l|}{CVE-2022-29854} &
  \multicolumn{1}{l|}{CVE-2022-29900} &
  CVE-2022-29901 \\ \hline
\multicolumn{1}{|l|}{CVE-2022-29945} &
  \multicolumn{1}{l|}{CVE-2022-29948} &
  \multicolumn{1}{l|}{CVE-2022-30111} &
  \multicolumn{1}{l|}{CVE-2022-31207} &
  \multicolumn{1}{l|}{CVE-2022-32503} &
  \multicolumn{1}{l|}{CVE-2022-32967} &
  \multicolumn{1}{l|}{CVE-2022-33196} &
  \multicolumn{1}{l|}{CVE-2022-34144} &
  \multicolumn{1}{l|}{CVE-2022-34634} &
  CVE-2022-34635 \\ \hline
\multicolumn{1}{|l|}{CVE-2022-34636} &
  \multicolumn{1}{l|}{CVE-2022-34641} &
  \multicolumn{1}{l|}{CVE-2022-35860} &
  \multicolumn{1}{l|}{CVE-2022-36307} &
  \multicolumn{1}{l|}{CVE-2022-36443} &
  \multicolumn{1}{l|}{CVE-2022-36945} &
  \multicolumn{1}{l|}{CVE-2022-37305} &
  \multicolumn{1}{l|}{CVE-2022-37418} &
  \multicolumn{1}{l|}{CVE-2022-37930} &
  CVE-2022-39902 \\ \hline
\multicolumn{1}{|l|}{CVE-2022-40633} &
  \multicolumn{1}{l|}{CVE-2022-40982} &
  \multicolumn{1}{l|}{CVE-2022-41505} &
  \multicolumn{1}{l|}{CVE-2022-41804} &
  \multicolumn{1}{l|}{CVE-2022-43096} &
  \multicolumn{1}{l|}{CVE-2022-45163} &
  \multicolumn{1}{l|}{CVE-2022-45190} &
  \multicolumn{1}{l|}{CVE-2022-45191} &
  \multicolumn{1}{l|}{CVE-2022-45192} &
  CVE-2022-45480 \\ \hline
\multicolumn{1}{|l|}{CVE-2022-47100} &
  \multicolumn{1}{l|}{CVE-2022-48251} &
  \multicolumn{1}{l|}{CVE-2022-48352} &
  \multicolumn{1}{l|}{CVE-2023-1526} &
  \multicolumn{1}{l|}{CVE-2023-20533} &
  \multicolumn{1}{l|}{CVE-2023-20569} &
  \multicolumn{1}{l|}{CVE-2023-20575} &
  \multicolumn{1}{l|}{CVE-2023-20579} &
  \multicolumn{1}{l|}{CVE-2023-20581} &
  CVE-2023-20582 \\ \hline
\multicolumn{1}{|l|}{CVE-2023-22655} &
  \multicolumn{1}{l|}{CVE-2023-23908} &
  \multicolumn{1}{l|}{CVE-2023-24033} &
  \multicolumn{1}{l|}{CVE-2023-25518} &
  \multicolumn{1}{l|}{CVE-2023-25758} &
  \multicolumn{1}{l|}{CVE-2023-26075} &
  \multicolumn{1}{l|}{CVE-2023-26076} &
  \multicolumn{1}{l|}{CVE-2023-26496} &
  \multicolumn{1}{l|}{CVE-2023-26941} &
  CVE-2023-26942 \\ \hline
\multicolumn{1}{|l|}{CVE-2023-28613} &
  \multicolumn{1}{l|}{CVE-2023-28896} &
  \multicolumn{1}{l|}{CVE-2023-30560} &
  \multicolumn{1}{l|}{CVE-2023-31114} &
  \multicolumn{1}{l|}{CVE-2023-31315} &
  \multicolumn{1}{l|}{CVE-2023-31759} &
  \multicolumn{1}{l|}{CVE-2023-32877} &
  \multicolumn{1}{l|}{CVE-2023-32878} &
  \multicolumn{1}{l|}{CVE-2023-32879} &
  CVE-2023-32880 \\ \hline
\multicolumn{1}{|l|}{CVE-2023-32881} &
  \multicolumn{1}{l|}{CVE-2023-32882} &
  \multicolumn{1}{l|}{CVE-2023-33037} &
  \multicolumn{1}{l|}{CVE-2023-33043} &
  \multicolumn{1}{l|}{CVE-2023-33281} &
  \multicolumn{1}{l|}{CVE-2023-34403} &
  \multicolumn{1}{l|}{CVE-2023-34625} &
  \multicolumn{1}{l|}{CVE-2023-3470} &
  \multicolumn{1}{l|}{CVE-2023-34724} &
  CVE-2023-35699 \\ \hline
\multicolumn{1}{|l|}{CVE-2023-36481} &
  \multicolumn{1}{l|}{CVE-2023-37195} &
  \multicolumn{1}{l|}{CVE-2023-37367} &
  \multicolumn{1}{l|}{CVE-2023-39843} &
  \multicolumn{1}{l|}{CVE-2023-40039} &
  \multicolumn{1}{l|}{CVE-2023-42482} &
  \multicolumn{1}{l|}{CVE-2023-42483} &
  \multicolumn{1}{l|}{CVE-2023-43122} &
  \multicolumn{1}{l|}{CVE-2023-43490} &
  CVE-2023-45864 \\ \hline
\multicolumn{1}{|l|}{CVE-2023-46033} &
  \multicolumn{1}{l|}{CVE-2023-47165} &
  \multicolumn{1}{l|}{CVE-2023-47262} &
  \multicolumn{1}{l|}{CVE-2023-47304} &
  \multicolumn{1}{l|}{CVE-2023-47616} &
  \multicolumn{1}{l|}{CVE-2023-48010} &
  \multicolumn{1}{l|}{CVE-2023-48034} &
  \multicolumn{1}{l|}{CVE-2023-49914} &
  \multicolumn{1}{l|}{CVE-2023-50126} &
  CVE-2023-50128 \\ \hline
\multicolumn{1}{|l|}{CVE-2023-50129} &
  \multicolumn{1}{l|}{CVE-2023-50430} &
  \multicolumn{1}{l|}{CVE-2023-50559} &
  \multicolumn{1}{l|}{CVE-2023-50805} &
  \multicolumn{1}{l|}{CVE-2023-50806} &
  \multicolumn{1}{l|}{CVE-2023-6068} &
  \multicolumn{1}{l|}{CVE-2023-7033} &
  \multicolumn{1}{l|}{CVE-2024-0160} &
  \multicolumn{1}{l|}{CVE-2024-1480} &
  CVE-2024-20034 \\ \hline
\multicolumn{1}{|l|}{CVE-2024-20889} &
  \multicolumn{1}{l|}{CVE-2024-20890} &
  \multicolumn{1}{l|}{CVE-2024-21478} &
  \multicolumn{1}{l|}{CVE-2024-21739} &
  \multicolumn{1}{l|}{CVE-2024-21739} &
  \multicolumn{1}{l|}{CVE-2024-21740} &
  \multicolumn{1}{l|}{CVE-2024-21741} &
  \multicolumn{1}{l|}{CVE-2024-21916} &
  \multicolumn{1}{l|}{CVE-2024-21981} &
  CVE-2024-22028 \\ \hline
\multicolumn{1}{|l|}{CVE-2024-22185} &
  \multicolumn{1}{l|}{CVE-2024-22247} &
  \multicolumn{1}{l|}{CVE-2024-22374} &
  \multicolumn{1}{l|}{CVE-2024-23360} &
  \multicolumn{1}{l|}{CVE-2024-23592} &
  \multicolumn{1}{l|}{CVE-2024-23806} &
  \multicolumn{1}{l|}{CVE-2024-23984} &
  \multicolumn{1}{l|}{CVE-2024-24051} &
  \multicolumn{1}{l|}{CVE-2024-24580} &
  CVE-2024-24853 \\ \hline
\multicolumn{1}{|l|}{CVE-2024-24980} &
  \multicolumn{1}{l|}{CVE-2024-24985} &
  \multicolumn{1}{l|}{CVE-2024-27361} &
  \multicolumn{1}{l|}{CVE-2024-27362} &
  \multicolumn{1}{l|}{CVE-2024-31798} &
  \multicolumn{1}{l|}{CVE-2024-32504} &
  \multicolumn{1}{l|}{CVE-2024-33038} &
  \multicolumn{1}{l|}{CVE-2024-33687} &
  \multicolumn{1}{l|}{CVE-2024-47667} &
  CVE-2024-4781 \\ \hline
\multicolumn{1}{|l|}{CVE-2024-4782} &
  \multicolumn{1}{l|}{CVE-2024-47968} &
  \multicolumn{1}{l|}{CVE-2024-47973} &
  \multicolumn{1}{l|}{CVE-2024-47974} &
  \multicolumn{1}{l|}{CVE-2024-48272} &
  \multicolumn{1}{l|}{CVE-2024-48796} &
  \multicolumn{1}{l|}{CVE-2024-48883} &
  \multicolumn{1}{l|}{CVE-2024-50920} &
  \multicolumn{1}{l|}{CVE-2024-50928} &
  CVE-2024-51529 \\ \hline
\multicolumn{1}{|l|}{CVE-2024-5209} &
  \multicolumn{1}{l|}{CVE-2024-5210} &
  \multicolumn{1}{l|}{CVE-2024-53832} &
  \multicolumn{1}{l|}{CVE-2024-54127} &
  \multicolumn{1}{l|}{CVE-2024-56161} &
  \multicolumn{1}{l|}{CVE-2024-6004} &
  \multicolumn{1}{l|}{CVE-2024-6242} &
  \multicolumn{1}{l|}{CVE-2024-6657} &
  \multicolumn{1}{l|}{CVE-2024-7726} &
  CVE-2024-7883 \\ \hline
\multicolumn{1}{|l|}{CVE-2024-9124} &
  \multicolumn{1}{l|}{CVE-2024-9834} &
  \multicolumn{1}{l|}{} &
  \multicolumn{1}{l|}{} &
  \multicolumn{1}{l|}{} &
  \multicolumn{1}{l|}{} &
  \multicolumn{1}{l|}{} &
  \multicolumn{1}{l|}{} &
  \multicolumn{1}{l|}{} &
   \\ \hline
\end{tabular}%
}
\label{tab:MIHW_List}
\end{table*}

\end{document}